# Cognitive Threat Intelligence and Explainable Federated Security Analytics for distributed Infrastructure Systems

**Md. Arifur Rahman[1]*, B. M. Taslimul Haque[2], Md. Iqbal Hossan[3], Md. Serajul Kabir Chowdhury Rubel[4]**

[1]Dept. of Information Studies, Trine University, Angola, Indiana, [2]Dept. of Business Information Systems, Central Michigan University, Mount Pleasant, Michigan, [3/4]Dept. of CS, Maharishi International University, Fairfield, Iowa, USA

## ABSTRACT

The increasing adoption of distributed infrastructure systems, cloud computing, Internet of Things (IoT) technologies, and edge-based architectures has significantly expanded the cybersecurity attack surface and introduced increasingly sophisticated cyber threats. Conventional centralized intrusion detection approaches often face challenges related to scalability, data privacy, communication overhead, and limited transparency in artificial intelligence-driven decision-making processes. To address these limitations, this study proposes a Cognitive Threat Intelligence and Explainable Federated Security Analytics framework for distributed infrastructure systems.

The proposed framework integrates Federated Learning (FL), Explainable Artificial Intelligence (XAI), and cognitive cybersecurity analytics to enable collaborative and privacy-preserving cyber threat detection across distributed network environments. Instead of transmitting sensitive raw network traffic data to centralized servers, local security models are independently trained at distributed nodes, where only encrypted model parameters and updates are shared through a federated aggregation mechanism. This decentralized learning architecture improves privacy protection while reducing communication dependency and centralized security risks.

To enhance intelligent threat analysis, the framework incorporates machine learning and deep learning algorithms including Random Forest, XGBoost, Autoencoder, and Long Short-Term Memory (LSTM) networks for anomaly detection and cyberattack classification. In addition, Explainable AI techniques such as SHAP and LIME are integrated to generate interpretable insights into anomaly predictions, enabling cybersecurity professionals to better understand the reasoning behind attack identification and risk assessment processes.

The effectiveness of the proposed framework is evaluated using benchmark cybersecurity datasets including NSL-KDD and CIC-IDS2017. Performance assessment is conducted using metrics such as accuracy, precision, recall, F1-score, ROC-AUC, detection latency, and communication efficiency. The expected outcomes of this research include improved intrusion detection capability, enhanced privacy preservation, reduced reliance on centralized infrastructures, and increased trustworthiness of AI-based cybersecurity systems. This study contributes to the development of intelligent, explainable, and resilient cybersecurity architectures designed to secure modern distributed infrastructure environments and critical digital systems.

## 1. INTRODUCTION

The rapid growth of distributed infrastructure systems, cloud computing platforms, Internet of Things (IoT) devices, edge computing technologies, and large-scale digital communication networks has transformed the modern cybersecurity landscape. Organizations increasingly rely on interconnected and decentralized infrastructures to support critical services, real-time communication, industrial automation, financial operations, healthcare systems, and intelligent digital services. Although these technologies improve operational efficiency, scalability, and connectivity, they also introduce significant cybersecurity vulnerabilities and expand the attack surface for sophisticated cyber threats.

Traditional cybersecurity mechanisms, particularly centralized intrusion detection systems, often struggle to effectively secure distributed network environments due to limitations associated with scalability, communication overhead, delayed threat response, and privacy concerns. Centralized architectures typically require sensitive network traffic and security logs to be transferred to a central server for analysis, which may increase the risk of data exposure, single points of failure, and compliance challenges in privacy-sensitive environments. In addition, many machine learning-based cybersecurity solutions operate as "black-box" systems, making it difficult for cybersecurity analysts and decision-makers to understand how anomaly predictions and threat classifications are generated.

Recent advancements in Artificial Intelligence (AI), Machine Learning (ML), Federated Learning (FL), and Explainable Artificial Intelligence (XAI) have introduced new opportunities for building intelligent and privacy-preserving cybersecurity frameworks. Federated Learning enables distributed devices and network nodes to collaboratively train cybersecurity models without sharing raw sensitive data, thereby preserving privacy while maintaining collective learning capabilities. At the same time, Explainable AI techniques improve the transparency and interpretability of AI-driven cybersecurity decisions by providing human-understandable explanations for anomaly detection and attack classification processes.

The integration of cognitive threat intelligence, federated analytics, and explainable anomaly detection represents a promising approach for addressing modern cybersecurity challenges in distributed infrastructure systems. By combining decentralized collaborative learning with interpretable AI-based security analytics, organizations can improve cyber threat detection accuracy, enhance trust in automated decision-making, and strengthen resilience against evolving cyberattacks. Therefore, there is a growing need for intelligent cybersecurity architectures that support scalable, explainable, and privacy-aware threat detection across modern distributed network environments and critical digital infrastructure systems.

## 2. PROBLEM STATEMENT

The increasing reliance on distributed infrastructure systems, cloud computing environments, Internet of Things (IoT) devices, and edge-based architectures has created complex cybersecurity challenges that traditional security mechanisms struggle to address effectively. Modern cyberattacks are becoming more sophisticated, adaptive, and difficult to detect due to the large volume of interconnected devices, decentralized communication patterns, and continuously evolving attack techniques. Conventional centralized intrusion detection systems often experience limitations related to scalability, communication overhead, delayed response time, and dependence on centralized data collection processes.

One of the major challenges in centralized cybersecurity systems is the requirement to transfer sensitive network traffic and security-related data to a central server for analysis. This approach increases the risk of data exposure, privacy violations, and single points of failure, particularly in critical infrastructure sectors such as healthcare, finance, transportation, and industrial control systems. In addition, distributed environments generate massive amounts of heterogeneous data, making centralized threat monitoring computationally expensive and inefficient.

Another significant problem is the lack of transparency in many Artificial Intelligence and Machine Learning-based cybersecurity models. Although advanced AI algorithms can achieve high anomaly detection accuracy, many operate as black-box systems that provide limited interpretability regarding how security decisions are made. This lack of explainability reduces trust among cybersecurity analysts and may create difficulties in incident investigation, compliance auditing, and operational decision-making.

Furthermore, existing cybersecurity frameworks often fail to simultaneously achieve privacy preservation, collaborative threat intelligence sharing, scalability, detection accuracy, and explainable decision-making within distributed network environments. As cyber threats continue to evolve across modern digital infrastructures, there is a critical need for an intelligent, decentralized, privacy-preserving, and interpretable cybersecurity framework capable of detecting anomalies and cyberattacks efficiently without compromising sensitive data. Therefore, this study addresses the need for a Cognitive Threat Intelligence and Explainable Federated Security Analytics framework that integrates Federated Learning and Explainable Artificial Intelligence to enhance cyber threat detection, improve privacy protection, and strengthen trust in AI-driven cybersecurity systems for distributed infrastructure environments.

## 3. OBJECTIVES OF THE STUDY

The main objective of this study is to develop a Cognitive Threat Intelligence and Explainable Federated Security Analytics framework for distributed infrastructure systems. The study aims to improve cyber threat detection accuracy, preserve sensitive data privacy, and enhance the transparency of AI-driven cybersecurity decision-making processes. This research also focuses on designing a federated learning-based cybersecurity framework for distributed network environments, integrating machine learning and deep learning techniques for anomaly detection, and incorporating Explainable Artificial Intelligence methods to provide interpretable threat analysis. In addition, the study evaluates the effectiveness of the proposed framework using benchmark cybersecurity datasets and standard performance evaluation metrics.To evaluate the proposed framework on the basis of performance metrics like accuracy, precision, recall, F1-score, and ROC-AUC.

## 4. RESEARCH QUESTIONS

This study seeks to address several important research questions related to cybersecurity threat detection in distributed infrastructure systems. The primary research questions are focused on how Federated Learning and Explainable Artificial Intelligence can improve cybersecurity performance, privacy preservation, and interpretability in distributed network environments. The study investigates whether a federated learning-based framework can effectively detect cyber threats without sharing sensitive raw network traffic data. It also examines how machine learning and deep learning algorithms contribute to anomaly detection accuracy in distributed cybersecurity systems. Furthermore, the research explores how Explainable Artificial Intelligence techniques such as SHAP and LIME can improve the transparency and interpretability of AI-driven cybersecurity decisions. Finally, the study evaluates whether the proposed framework can enhance communication efficiency, scalability, and overall trustworthiness in modern distributed infrastructure security environments.

## 5. SIGNIFICANCE OF THE STUDY

This study is significant because it addresses critical cybersecurity challenges associated with distributed infrastructure systems, cloud computing, Internet of Things (IoT) environments, and edge-based networks. As cyber threats continue to evolve in complexity and scale, there is an increasing need for intelligent cybersecurity frameworks that can provide accurate threat detection while preserving data privacy and ensuring transparency in decision-making processes.

The proposed Cognitive Threat Intelligence and Explainable Federated Security Analytics framework contribute to the advancement of privacy-preserving and decentralized cybersecurity solutions by integrating Federated Learning and Explainable Artificial Intelligence techniques. The study helps reduce reliance on centralized security architectures, minimize sensitive data exposure, and improve collaborative threat intelligence across distributed network environments.

In addition, the integration of explainable AI techniques enhances trust and interpretability in AI-driven cybersecurity systems by enabling security analysts to better understand anomaly detection and attack classification processes. The findings of this research may support organizations, researchers, and cybersecurity professionals in developing scalable, transparent, and resilient cybersecurity solutions for protecting modern digital infrastructures and critical network systems.

## 6. LITERATURE REVIEW

A. Cognitive Threat Intelligence and Explainable Federated Security Analytics

The rapid advancement of distributed infrastructure systems, cloud computing, Internet of Things (IoT) technologies, and edge computing environments has significantly transformed modern cybersecurity operations. As organizations increasingly depend on decentralized digital infrastructures, cyber threats have become more sophisticated, dynamic, and difficult to detect using traditional security mechanisms. Researchers have explored various Artificial Intelligence (AI), Machine Learning (ML), Federated Learning (FL), and Explainable Artificial Intelligence (XAI) approaches to improve cyber threat detection, anomaly analysis, and network security resilience.

Traditional intrusion detection systems primarily rely on centralized architectures where network traffic and security data are collected and analysed at a central location. Although centralized systems can achieve effective monitoring in smaller environments, they often face challenges related to scalability, communication overhead, delayed response, and privacy risks in large distributed networks. Several studies have highlighted that centralized cybersecurity frameworks may expose sensitive data to unauthorized access and create single points of failure in critical infrastructure systems.

Machine learning and deep learning techniques have become widely adopted in cybersecurity research for detecting malicious activities and identifying network anomalies. Algorithms such as Random Forest, Support Vector Machine (SVM), XGBoost, Autoencoder, Convolutional Neural Networks (CNN), and Long Short-Term Memory (LSTM) networks have demonstrated strong performance in intrusion detection and attack classification tasks. Deep learning-based models are particularly effective in recognizing complex patterns within large-scale network traffic datasets. However, many AI-driven cybersecurity systems operate as black-box models, limiting transparency and reducing trust in automated decision-making processes.

Federated Learning has recently emerged as a promising decentralized learning approach for privacy-preserving cybersecurity applications. Unlike traditional centralized machine learning methods, Federated Learning enables multiple distributed devices or network nodes to collaboratively train models without sharing raw sensitive data. Instead, local model updates are aggregated through a centralized or distributed coordination mechanism. Existing studies suggest that Federated Learning can improve privacy protection, reduce data exposure risks, and support collaborative threat intelligence across distributed environments. Nevertheless, challenges related to communication efficiency, model heterogeneity, scalability, and adversarial attacks remain active areas of research.

Explainable Artificial Intelligence techniques such as SHAP (Shapley Additive Explanations) and LIME (Local Interpretable Model-Agnostic Explanations) have been increasingly integrated into cybersecurity systems to improve interpretability and transparency. These techniques help security analysts understand how AI models generate anomaly predictions and attack classifications. Explainability is particularly important in critical infrastructure environments where cybersecurity decisions must be interpretable, trustworthy, and compliant with organizational and regulatory requirements.

Although previous studies have separately explored Federated Learning, anomaly detection, and Explainable AI in cybersecurity, limited research has fully integrated cognitive threat intelligence, explainable analytics, and privacy-preserving federated learning within distributed infrastructure systems. Therefore, this study contributes to the existing literature by proposing a Cognitive Threat Intelligence and Explainable Federated Security Analytics framework that combines decentralized collaborative learning, intelligent anomaly detection, and interpretable AI-driven cybersecurity analysis for modern distributed network environments.

A. Explainable Federated Security Frameworks

Explainable Federated Security Frameworks is a concise and research-oriented title that emphasizes the integration of Explainable Artificial Intelligence (XAI) and Federated

Learning within cybersecurity systems. The title reflects the study's focus on privacy-preserving collaborative security analytics, transparent anomaly detection, and intelligent threat analysis in distributed infrastructure environments. It highlights the development of scalable and interpretable cybersecurity frameworks designed to improve trust, resilience, and security performance in modern distributed network systems

Empirical Study

An empirical study is a research approach based on real-world observation, experimentation, data collection, and practical analysis rather than only theoretical discussion. In cybersecurity and artificial intelligence research, empirical studies are conducted to evaluate the effectiveness, accuracy, and performance of proposed models or frameworks using actual datasets, experiments, simulations, or measurable results.

In this study, the empirical approach is applied to evaluate the proposed Explainable Federated Security Framework using benchmark cybersecurity datasets such as NSL-KDD and CIC-IDS2017. Machine learning and deep learning algorithms including Random Forest, XGBoost, Autoencoder, and LSTM networks are experimentally analyzed for anomaly detection and cyber threat classification. The framework performance is measured using evaluation metrics such as accuracy, precision, recall, F1-score, ROC-AUC, detection latency, and communication efficiency. The empirical analysis helps determine the effectiveness of the proposed framework in improving cybersecurity threat detection, privacy preservation, and explainability within distributed infrastructure systems.

## 7. METHODOLOGY

This study adopts an empirical and experimental research methodology to develop and evaluate the proposed Explainable Federated Security Framework for distributed infrastructure systems. The methodology integrates Federated Learning (FL), machine learning, deep learning, and Explainable Artificial Intelligence (XAI) techniques for privacy-preserving cyber threat detection and anomaly analysis.

The research utilizes benchmark cybersecurity datasets including NSL-KDD and CIC-IDS2017 to analyze malicious network activities and intrusion patterns within distributed environments. Data preprocessing techniques such as normalization, feature selection, data cleaning, and traffic classification are applied to improve data quality and model performance.

The proposed framework uses distributed network nodes where local anomaly detection models are independently trained without sharing raw sensitive network traffic data. Instead, model parameters and updates are exchanged through a federated aggregation mechanism to preserve privacy and support collaborative learning. Machine learning and deep learning algorithms including Random Forest, XGBoost, Autoencoder, and Long Short-Term Memory (LSTM) networks are implemented for cyber threat detection and anomaly classification.

To improve transparency and interpretability, Explainable Artificial Intelligence techniques such as SHAP and LIME are integrated into the framework to explain anomaly predictions and attack classifications. The performance of the proposed system is evaluated using metrics including accuracy, precision, recall, F1-score, ROC-AUC, detection latency, and communication efficiency. The experimental results are comparatively analysed to determine the effectiveness of the framework in improving distributed cybersecurity intelligence, privacy preservation, and explainable threat detection.

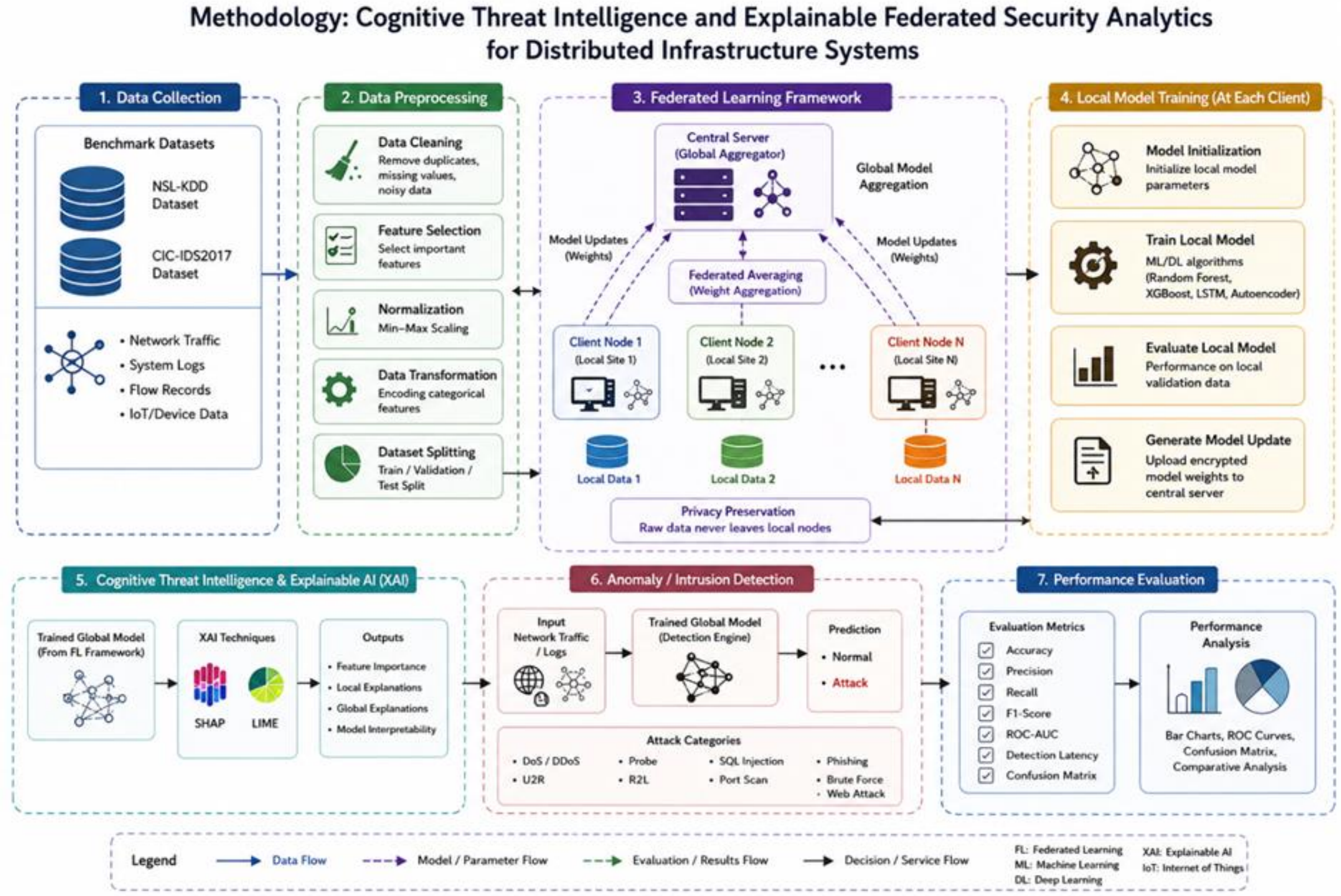

This workflow illustrates Cognitive Threat Intelligence and Explainable Federated Security Analytics for Distributed Infrastructure Systems

The methodology diagram represents the operational workflow of the proposed Cognitive Threat Intelligence and Explainable Federated Security Analytics framework for distributed infrastructure systems. The framework collects and preprocesses cybersecurity datasets before applying federated learning to train distributed machine learning and deep learning models such as XGBoost, LSTM, Random Forest, and Autoencoder at local client nodes. Local model updates are aggregated through Federated Averaging to preserve privacy without sharing raw data. Explainable AI techniques including SHAP and LIME are integrated to improve transparency and interpretability in anomaly detection decisions. Finally, the framework performs intrusion detection and evaluates system performance using cybersecurity metrics such as accuracy, precision, recall, F1-score, ROC-AUC, and confusion matrix analysis.

**Research Design**

This study adopts an empirical and experimental research design to develop and evaluate the proposed Cognitive Threat Intelligence and Explainable Federated Security Analytics framework. The research integrates Federated Learning, machine learning, deep learning, and Explainable Artificial Intelligence techniques for privacy-preserving anomaly detection in distributed infrastructure systems. Benchmark datasets such as NSL-KDD and CIC-IDS2017 are used to evaluate system performance based on accuracy, precision, recall, F1-score, and ROC-AUC

**Dataset Selection and Data Collection**

This study utilizes benchmark cybersecurity datasets including NSL-KDD and CIC-IDS2017 for anomaly detection and cyber threat analysis in distributed infrastructure environments. These datasets contain both normal and malicious network traffic records representing different types of cyberattacks such as DoS, DDoS, Probe, R2L, U2R, brute force, and web attacks.

Data collection focuses on network traffic information, system logs, flow records, and intrusion-related features relevant to distributed cybersecurity environments. The selected datasets are widely used in cybersecurity research because they provide reliable and standardized data for evaluating machine learning, deep learning, and federated learning-based intrusion detection systems.

NSL-KDD Dataset

| 1 | duration | protocol_type | service | flag | src_bytes | dst_bytes | land | wrong_fragment | urgent | hot | num_failed_logins | logged_in | num_compromised | root_shell | su_attempted | num_r… | num_file_creat… | num_sh… | num_access_fil… | num_outbound… | is_host_login | is_guest_login |
|---|---|---|---|---|---|---|---|---|---|---|---|---|---|---|---|---|---|---|---|---|---|---|
| 2 | 0 | tcp | private | REJ | 0 | 0 | 0 | 0 | 0 | 0 | 0 | 0 | 0 | 0 | 0 | 0 | 0 | 0 | 0 | 0 | 0 | 0 |
| 3 | 0 | tcp | private | REJ | 0 | 0 | 0 | 0 | 0 | 0 | 0 | 0 | 0 | 0 | 0 | 0 | 0 | 0 | 0 | 0 | 0 | 0 |
| 4 | 2 | tcp | ftp_data | SF | 12983 | 0 | 0 | 0 | 0 | 0 | 0 | 0 | 0 | 0 | 0 | 0 | 0 | 0 | 0 | 0 | 0 | 0 |
| 5 | 0 | icmp | eco_i | SF | 20 | 0 | 0 | 0 | 0 | 0 | 0 | 0 | 0 | 0 | 0 | 0 | 0 | 0 | 0 | 0 | 0 | 0 |
| 6 | 1 | tcp | telnet | RSTO | 0 | 15 | 0 | 0 | 0 | 0 | 0 | 0 | 0 | 0 | 0 | 0 | 0 | 0 | 0 | 0 | 0 | 0 |
| 7 | 0 | tcp | http | SF | 267 | 14515 | 0 | 0 | 0 | 0 | 0 | 1 | 0 | 0 | 0 | 0 | 0 | 0 | 0 | 0 | 0 | 0 |
| 8 | 0 | tcp | smtp | SF | 1022 | 387 | 0 | 0 | 0 | 0 | 0 | 1 | 0 | 0 | 0 | 0 | 0 | 0 | 0 | 0 | 0 | 0 |
| 9 | 0 | tcp | telnet | SF | 129 | 174 | 0 | 0 | 0 | 0 | 1 | 0 | 0 | 0 | 0 | 0 | 0 | 0 | 0 | 0 | 0 | 0 |
| 10 | 0 | tcp | http | SF | 327 | 467 | 0 | 0 | 0 | 0 | 0 | 1 | 0 | 0 | 0 | 0 | 0 | 0 | 0 | 0 | 0 | 0 |
| 11 | 0 | tcp | ftp | SF | 26 | 157 | 0 | 0 | 0 | 0 | 1 | 0 | 0 | 0 | 0 | 0 | 0 | 0 | 0 | 0 | 0 | 1 |
| 12 | 0 | tcp | telnet | SF | 0 | 0 | 0 | 0 | 0 | 0 | 0 | 0 | 0 | 0 | 0 | 0 | 0 | 0 | 0 | 0 | 0 | 0 |
| 13 | 0 | tcp | smtp | SF | 616 | 330 | 0 | 0 | 0 | 0 | 0 | 1 | 0 | 0 | 0 | 0 | 0 | 0 | 0 | 0 | 0 | 0 |
| 14 | 0 | tcp | private | REJ | 0 | 0 | 0 | 0 | 0 | 0 | 0 | 0 | 0 | 0 | 0 | 0 | 0 | 0 | 0 | 0 | 0 | 0 |
| 15 | 0 | tcp | telnet | S0 | 0 | 0 | 0 | 0 | 0 | 0 | 0 | 0 | 0 | 0 | 0 | 0 | 0 | 0 | 0 | 0 | 0 | 0 |
| 16 | 37 | tcp | telnet | SF | 773 | 364200 | 0 | 0 | 0 | 0 | 0 | 1 | 0 | 0 | 0 | 0 | 0 | 0 | 0 | 0 | 0 | 0 |
| 17 | 0 | tcp | http | SF | 350 | 3610 | 0 | 0 | 0 | 0 | 0 | 1 | 0 | 0 | 0 | 0 | 0 | 0 | 0 | 0 | 0 | 0 |
| 18 | 0 | tcp | http | SF | 213 | 659 | 0 | 0 | 0 | 0 | 0 | 1 | 0 | 0 | 0 | 0 | 0 | 0 | 0 | 0 | 0 | 0 |
| 19 | 0 | tcp | http | SF | 246 | 2090 | 0 | 0 | 0 | 0 | 0 | 1 | 0 | 0 | 0 | 0 | 0 | 0 | 0 | 0 | 0 | 0 |
| 20 | 0 | udp | private | SF | 45 | 44 | 0 | 0 | 0 | 0 | 0 | 0 | 0 | 0 | 0 | 0 | 0 | 0 | 0 | 0 | 0 | 0 |
| 21 | 0 | tcp | private | REJ | 0 | 0 | 0 | 0 | 0 | 0 | 0 | 0 | 0 | 0 | 0 | 0 | 0 | 0 | 0 | 0 | 0 | 0 |
| 22 | 0 | tcp | ldap | REJ | 0 | 0 | 0 | 0 | 0 | 0 | 0 | 0 | 0 | 0 | 0 | 0 | 0 | 0 | 0 | 0 | 0 | 0 |
| 23 | 0 | tcp | pop_3 | S0 | 0 | 0 | 0 | 0 | 0 | 0 | 0 | 0 | 0 | 0 | 0 | 0 | 0 | 0 | 0 | 0 | 0 | 0 |
| 24 | 0 | tcp | http | SF | 196 | 1823 | 0 | 0 | 0 | 0 | 0 | 1 | 0 | 0 | 0 | 0 | 0 | 0 | 0 | 0 | 0 | 0 |
| 25 | 0 | tcp | http | SF | 277 | 1816 | 0 | 0 | 0 | 0 | 0 | 1 | 0 | 0 | 0 | 0 | 0 | 0 | 0 | 0 | 0 | 0 |
| 26 | 0 | tcp | courier | REJ | 0 | 0 | 0 | 0 | 0 | 0 | 0 | 0 | 0 | 0 | 0 | 0 | 0 | 0 | 0 | 0 | 0 | 0 |
| 27 | 0 | tcp | discard | RSTO | 0 | 0 | 0 | 0 | 0 | 0 | 0 | 0 | 0 | 0 | 0 | 0 | 0 | 0 | 0 | 0 | 0 | 0 |
| 28 | 0 | tcp | http | SF | 294 | 6442 | 0 | 0 | 0 | 0 | 0 | 1 | 0 | 0 | 0 | 0 | 0 | 0 | 0 | 0 | 0 | 0 |
| 29 | 0 | tcp | http | SF | 300 | 440 | 0 | 0 | 0 | 0 | 0 | 1 | 0 | 0 | 0 | 0 | 0 | 0 | 0 | 0 | 0 | 0 |
| 30 | 0 | icmp | ecr_i | SF | 520 | 0 | 0 | 0 | 0 | 0 | 0 | 0 | 0 | 0 | 0 | 0 | 0 | 0 | 0 | 0 | 0 | 0 |
| 31 | 0 | udp | private | SF | 54 | 51 | 0 | 0 | 0 | 0 | 0 | 0 | 0 | 0 | 0 | 0 | 0 | 0 | 0 | 0 | 0 | 0 |
| 32 | 805 | tcp | http | RSTR | 76944 | 1 | 0 | 0 | 0 | 0 | 0 | 1 | 0 | 0 | 0 | 0 | 0 | 0 | 0 | 0 | 0 | 0 |
| 33 | 0 | tcp | smtp | SF | 720 | 281 | 0 | 0 | 0 | 0 | 0 | 1 | 0 | 0 | 0 | 0 | 0 | 0 | 0 | 0 | 0 | 0 |
| 34 | 0 | tcp | http | SF | 301 | 19794 | 0 | 0 | 0 | 0 | 0 | 1 | 0 | 0 | 0 | 0 | 0 | 0 | 0 | 0 | 0 | 0 |
| 35 | 0 | udp | private | SF | 1 | 1 | 0 | 0 | 0 | 0 | 0 | 0 | 0 | 0 | 0 | 0 | 0 | 0 | 0 | 0 | 0 | 0 |

| 1 | count | srv_count | serror_rate | srv_serror_rate | rerror_rate | srv_rerror_rate | same_srv_rate | diff_srv_rate | srv_diff_host_rate | dst_host_count | dst_host_srv_count | dst_host_same_srv… | dst_host_diff_srv_rate | dst_host_same_src_port_rate | dst_host_srv_diff_host_rate | dst_host_serror_rate | dst_host_srv_serror_rate | dst_host_rerror_rate | dst_host_srv_rerror_rate | class | classnum |
|---|---|---|---|---|---|---|---|---|---|---|---|---|---|---|---|---|---|---|---|---|---|
| 2 | 229 | 10 | 0 | 0 | 1 | 1 | 0.04 | 0.06 | 0 | 255 | 10 | 0 | 0.06 | 0 | 0 | 0 | 0 | 1 | 1 | neptune | 21 |
| 3 | 136 | 1 | 0 | 0 | 1 | 1 | 0.01 | 0.06 | 0 | 255 | 1 | 0 | 0.06 | 0 | 0 | 0 | 0 | 1 | 1 | neptune | 21 |
| 4 | 1 | 1 | 0 | 0 | 0 | 0 | 1 | 0 | 0 | 134 | 86 | 0.6 | 0.04 | 0.61 | 0.02 | 0 | 0 | 0 | 0 | normal | 21 |
| 5 | 1 | 65 | 0 | 0 | 0 | 0 | 1 | 0 | 1 | 3 | 57 | 1 | 0 | 1 | 0.28 | 0 | 0 | 0 | 0 | saint | 15 |
| 6 | 1 | 8 | 0 | 0.1 | 1 | 0.5 | 1 | 0 | 0.75 | 29 | 86 | 0.3 | 0.17 | 0.03 | 0.02 | 0 | 0 | 0.83 | 0.71 | mscan | 11 |
| 7 | 4 | 4 | 0 | 0 | 0 | 0 | 1 | 0 | 0 | 155 | 255 | 1 | 0 | 0.01 | 0.03 | 0.01 | 0 | 0 | 0 | normal | 21 |
| 8 | 1 | 3 | 0 | 0 | 0 | 0 | 1 | 0 | 1 | 255 | 28 | 0.1 | 0.72 | 0 | 0 | 0 | 0 | 0.72 | 0.04 | normal | 21 |
| 9 | 1 | 1 | 0 | 0 | 0 | 0 | 1 | 0 | 0 | 255 | 255 | 1 | 0 | 0 | 0 | 0.01 | 0.01 | 0.02 | 0.02 | guess_pas | 15 |
| 10 | 33 | 47 | 0 | 0 | 0 | 0 | 1 | 0 | 0.04 | 151 | 255 | 1 | 0 | 0.01 | 0.03 | 0 | 0 | 0 | 0 | normal | 21 |
| 11 | 1 | 1 | 0 | 0 | 0 | 0 | 1 | 0 | 0 | 52 | 26 | 0.5 | 0.08 | 0.02 | 0 | 0 | 0 | 0 | 0 | guess_pas | 7 |
| 12 | 1 | 1 | 0 | 0 | 0 | 0 | 1 | 0 | 0 | 255 | 128 | 0.5 | 0.01 | 0 | 0 | 0 | 0 | 0.66 | 0.32 | mscan | 9 |
| 13 | 1 | 2 | 0 | 0 | 0 | 0 | 1 | 0 | 1 | 255 | 129 | 0.5 | 0.03 | 0 | 0 | 0 | 0 | 0.33 | 0 | normal | 18 |
| 14 | 111 | 2 | 0 | 0 | 1 | 1 | 0.02 | 0.07 | 0 | 255 | 2 | 0 | 0.07 | 0 | 0 | 0 | 0 | 1 | 1 | neptune | 21 |
| 15 | 120 | 120 | 1 | 1 | 0 | 0 | 1 | 0 | 0 | 235 | 171 | 0.7 | 0.07 | 0 | 0 | 0.69 | 0.95 | 0.02 | 0 | neptune | 18 |
| 16 | 1 | 1 | 0 | 0 | 0 | 0 | 1 | 0 | 0 | 38 | 73 | 0.2 | 0.05 | 0.03 | 0.04 | 0 | 0.77 | 0 | 0.07 | normal | 14 |
| 17 | 8 | 8 | 0 | 0 | 0 | 0 | 1 | 0 | 0 | 71 | 255 | 1 | 0 | 0.01 | 0.04 | 0 | 0 | 0 | 0 | normal | 21 |
| 18 | 24 | 24 | 0 | 0 | 0 | 0 | 1 | 0 | 0 | 255 | 255 | 1 | 0 | 0 | 0 | 0 | 0 | 0 | 0 | normal | 21 |
| 19 | 16 | 16 | 0 | 0 | 0 | 0 | 1 | 0 | 0 | 35 | 255 | 1 | 0 | 0.03 | 0.05 | 0 | 0 | 0 | 0 | normal | 21 |
| 20 | 505 | 505 | 0 | 0 | 0 | 0 | 1 | 0 | 0 | 255 | 255 | 1 | 0 | 1 | 0 | 0 | 0 | 0 | 0 | normal | 15 |
| 21 | 204 | 18 | 0 | 0 | 1 | 1 | 0.09 | 0.07 | 0 | 255 | 18 | 0.1 | 0.07 | 0 | 0 | 0 | 0 | 1 | 1 | neptune | 21 |
| 22 | 118 | 19 | 0 | 0 | 1 | 1 | 0.16 | 0.05 | 0 | 255 | 19 | 0.1 | 0.05 | 0 | 0 | 0 | 0 | 1 | 1 | neptune | 21 |
| 23 | 1 | 1 | 1 | 1 | 0 | 0 | 1 | 0 | 0 | 255 | 87 | 0.3 | 0.01 | 0.01 | 0 | 1 | 1 | 0 | 0 | mscan | 18 |
| 24 | 17 | 17 | 0 | 0 | 0 | 0 | 1 | 0 | 0 | 255 | 255 | 1 | 0 | 0 | 0 | 0 | 0 | 0 | 0 | normal | 21 |
| 25 | 17 | 18 | 0 | 0 | 0 | 0 | 1 | 0 | 0.11 | 36 | 255 | 1 | 0 | 0.03 | 0.02 | 0 | 0 | 0 | 0 | normal | 21 |
| 26 | 116 | 8 | 0 | 0 | 1 | 1 | 0.07 | 0.07 | 0 | 255 | 8 | 0 | 0.06 | 0 | 0 | 0 | 0 | 1 | 1 | neptune | 18 |
| 27 | 273 | 13 | 0 | 0 | 1 | 1 | 0.05 | 0.06 | 0 | 255 | 13 | 0.1 | 0.06 | 0 | 0 | 0 | 0 | 1 | 1 | neptune | 20 |
| 28 | 22 | 46 | 0 | 0 | 0 | 0 | 1 | 0 | 0.11 | 180 | 255 | 1 | 0 | 0.01 | 0.01 | 0 | 0 | 0 | 0 | normal | 21 |
| 29 | 7 | 7 | 0 | 0 | 0 | 0 | 1 | 0 | 0 | 255 | 255 | 1 | 0 | 0 | 0 | 0 | 0 | 0 | 0 | normal | 21 |
| 30 | 511 | 511 | 0 | 0 | 0 | 0 | 1 | 0 | 0 | 46 | 59 | 1 | 0 | 1 | 0.14 | 0 | 0 | 0 | 0 | smurf | 18 |
| 31 | 511 | 511 | 0 | 0 | 0 | 0 | 1 | 0 | 0 | 255 | 255 | 1 | 0 | 0.83 | 0 | 0 | 0 | 0 | 0 | normal | 17 |
| 32 | 12 | 12 | 0 | 0 | 1 | 1 | 1 | 0 | 0 | 241 | 238 | 1 | 0.01 | 0 | 0 | 0 | 0 | 0.07 | 0.07 | apache2 | 14 |
| 33 | 1 | 2 | 0 | 0 | 0 | 0 | 1 | 0 | 1 | 158 | 82 | 0.5 | 0.06 | 0.01 | 0 | 0.01 | 0 | 0 | 0 | normal | 21 |
| 34 | 15 | 15 | 0 | 0 | 0 | 0 | 1 | 0 | 0 | 20 | 255 | 1 | 0 | 0.05 | 0.02 | 0.05 | 0 | 0 | 0 | normal | 21 |
| 35 | 40 | 3 | 0 | 0 | 0 | 0 | 0.08 | 0.38 | 0 | 255 | 3 | 0 | 0.58 | 0.99 | 0 | 0 | 0 | 0.01 | 0 | normal | 1 |

KDDTest

(Source Link: https://www.kaggle.com/datasets/dhoogla/cicidscollection)

The NSL-KDD dataset has been chosen as the main dataset used for the research due to its well acceptance for intrusion detection and anomaly detection purposes. The dataset has been improved from the KDD99 dataset and cleaned up to eliminate redundant records that would have a negative impact on machine learning. It includes attack categories like Denial of Service (DoS), Probe, Remote-to-Local (R2L), User-to-Root (U2R) attacks [16]. The dataset provides information on network traffic characteristics such as protocol type, service, source bytes, destination bytes, duration, and flag status, which are ideal for explainable anomaly detection and federated learning experiments.

CIC-IDS2017 Dataset

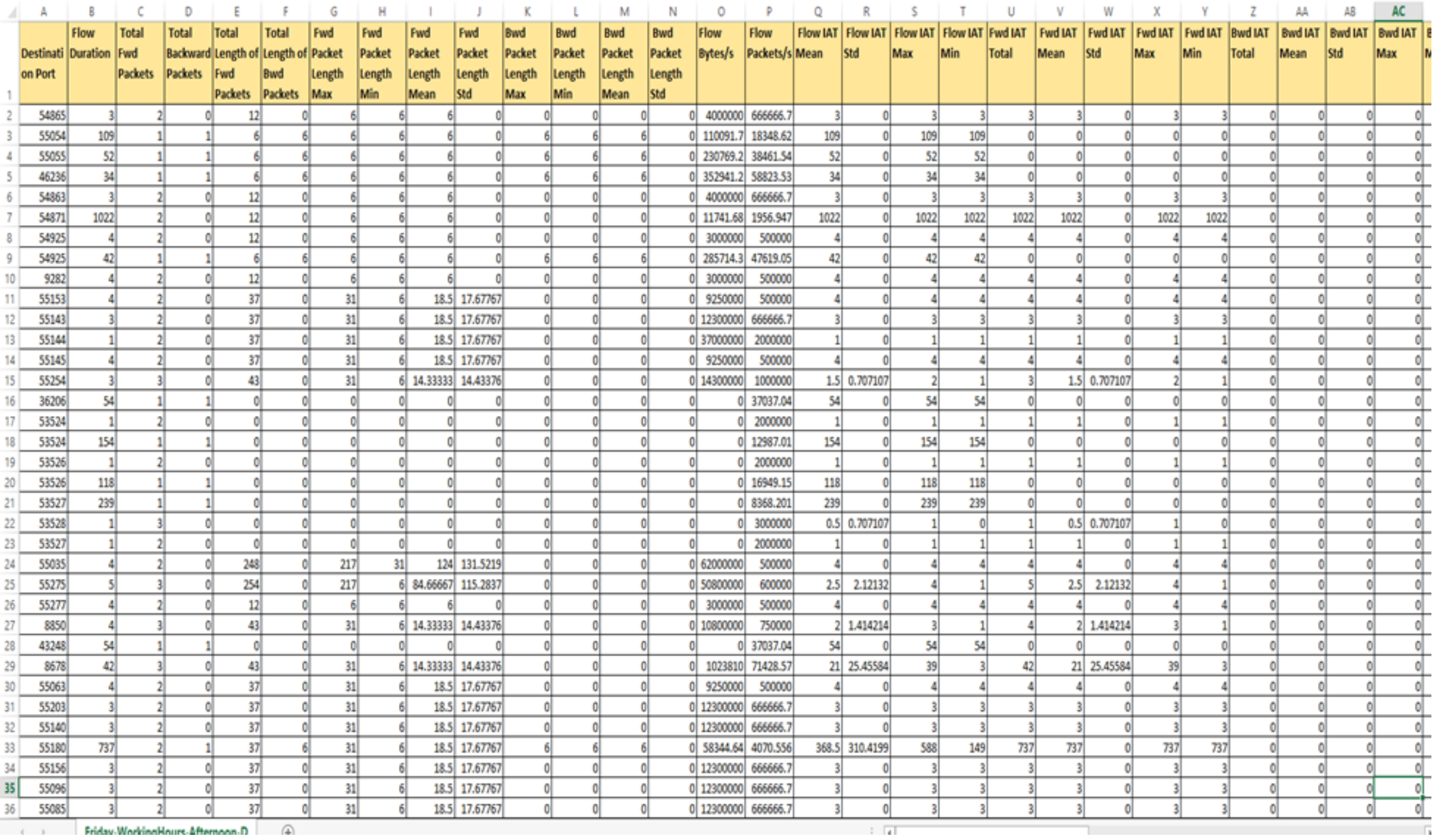

| | A | B | C | D | E | F | G | H | I | J | K | L | M | N |
|---|---|---|---|---|---|---|---|---|---|---|---|---|---|---|
| 1 | Destination Port | Flow Duration | Total Fwd Packets | Total Backward Packets | Total Length of Fwd Packets | Total Length of Bwd Packets | Fwd Packet Length Max | Fwd Packet Length Min | Fwd Packet Length Mean | Fwd Packet Length Std | Bwd Packet Length Max | Bwd Packet Length Min | Bwd Packet Length Mean | Bwd Packet Length Std |
| 2 | 54865 | 3 | 2 | 0 | 12 | 0 | 6 | 6 | 6 | 0 | 0 | 0 | 0 | 0 |
| 3 | 55054 | 109 | 1 | 1 | 6 | 6 | 6 | 6 | 6 | 0 | 6 | 6 | 6 | 0 |
| 4 | 55055 | 52 | 1 | 1 | 6 | 6 | 6 | 6 | 6 | 0 | 6 | 6 | 6 | 0 |
| 5 | 46236 | 34 | 1 | 1 | 6 | 6 | 6 | 6 | 6 | 0 | 6 | 6 | 6 | 0 |
| 6 | 54863 | 3 | 2 | 0 | 12 | 0 | 6 | 6 | 6 | 0 | 0 | 0 | 0 | 0 |
| 7 | 54871 | 1022 | 2 | 0 | 12 | 0 | 6 | 6 | 6 | 0 | 0 | 0 | 0 | 0 |
| 8 | 54925 | 4 | 2 | 0 | 12 | 0 | 6 | 6 | 6 | 0 | 0 | 0 | 0 | 0 |
| 9 | 54925 | 42 | 1 | 1 | 6 | 6 | 6 | 6 | 6 | 0 | 6 | 6 | 6 | 0 |
| 10 | 9282 | 4 | 2 | 0 | 12 | 0 | 6 | 6 | 6 | 0 | 0 | 0 | 0 | 0 |
| 11 | 55153 | 4 | 2 | 0 | 37 | 0 | 31 | 6 | 18.5 | 17.67767 | 0 | 0 | 0 | 0 |
| 12 | 55143 | 3 | 2 | 0 | 37 | 0 | 31 | 6 | 18.5 | 17.67767 | 0 | 0 | 0 | 0 |
| 13 | 55144 | 1 | 2 | 0 | 37 | 0 | 31 | 6 | 18.5 | 17.67767 | 0 | 0 | 0 | 0 |
| 14 | 55145 | 4 | 2 | 0 | 37 | 0 | 31 | 6 | 18.5 | 17.67767 | 0 | 0 | 0 | 0 |
| 15 | 55254 | 3 | 3 | 0 | 43 | 0 | 31 | 6 | 14.33333 | 14.43376 | 0 | 0 | 0 | 0 |
| 16 | 36206 | 54 | 1 | 1 | 0 | 0 | 0 | 0 | 0 | 0 | 0 | 0 | 0 | 0 |
| 17 | 53524 | 1 | 2 | 0 | 0 | 0 | 0 | 0 | 0 | 0 | 0 | 0 | 0 | 0 |
| 18 | 53524 | 154 | 1 | 1 | 0 | 0 | 0 | 0 | 0 | 0 | 0 | 0 | 0 | 0 |
| 19 | 53526 | 1 | 2 | 0 | 0 | 0 | 0 | 0 | 0 | 0 | 0 | 0 | 0 | 0 |
| 20 | 53526 | 118 | 1 | 1 | 0 | 0 | 0 | 0 | 0 | 0 | 0 | 0 | 0 | 0 |
| 21 | 53527 | 239 | 1 | 1 | 0 | 0 | 0 | 0 | 0 | 0 | 0 | 0 | 0 | 0 |
| 22 | 53528 | 1 | 3 | 0 | 0 | 0 | 0 | 0 | 0 | 0 | 0 | 0 | 0 | 0 |
| 23 | 53527 | 1 | 2 | 0 | 0 | 0 | 0 | 0 | 0 | 0 | 0 | 0 | 0 | 0 |
| 24 | 55035 | 4 | 2 | 0 | 248 | 0 | 217 | 31 | 124 | 131.5219 | 0 | 0 | 0 | 0 |
| 25 | 55275 | 5 | 3 | 0 | 254 | 0 | 217 | 6 | 84.66667 | 115.2837 | 0 | 0 | 0 | 0 |
| 26 | 55277 | 4 | 2 | 0 | 12 | 0 | 6 | 6 | 6 | 0 | 0 | 0 | 0 | 0 |
| 27 | 8850 | 4 | 3 | 0 | 43 | 0 | 31 | 6 | 14.33333 | 14.43376 | 0 | 0 | 0 | 0 |
| 28 | 43248 | 54 | 1 | 1 | 0 | 0 | 0 | 0 | 0 | 0 | 0 | 0 | 0 | 0 |
| 29 | 8678 | 42 | 3 | 0 | 43 | 0 | 31 | 6 | 14.33333 | 14.43376 | 0 | 0 | 0 | 0 |
| 30 | 55063 | 4 | 2 | 0 | 37 | 0 | 31 | 6 | 18.5 | 17.67767 | 0 | 0 | 0 | 0 |
| 31 | 55203 | 3 | 2 | 0 | 37 | 0 | 31 | 6 | 18.5 | 17.67767 | 0 | 0 | 0 | 0 |
| 32 | 55140 | 3 | 2 | 0 | 37 | 0 | 31 | 6 | 18.5 | 17.67767 | 0 | 0 | 0 | 0 |
| 33 | 55180 | 737 | 2 | 1 | 37 | 6 | 31 | 6 | 18.5 | 17.67767 | 6 | 6 | 6 | 0 |
| 34 | 55156 | 3 | 2 | 0 | 37 | 0 | 31 | 6 | 18.5 | 17.67767 | 0 | 0 | 0 | 0 |
| 35 | 55096 | 3 | 2 | 0 | 37 | 0 | 31 | 6 | 18.5 | 17.67767 | 0 | 0 | 0 | 0 |
| 36 | 55085 | 3 | 2 | 0 | 37 | 0 | 31 | 6 | 18.5 | 17.67767 | 0 | 0 | 0 | 0 |

| | O | P | Q | R | S | T | U | V | W | X | Y | Z | AA | AB | AC |
|---|---|---|---|---|---|---|---|---|---|---|---|---|---|---|---|
| 1 | Flow Bytes/s | Flow Packets/s | Flow IAT Mean | Flow IAT Std | Flow IAT Max | Flow IAT Min | Fwd IAT Total | Fwd IAT Mean | Fwd IAT Std | Fwd IAT Max | Fwd IAT Min | Bwd IAT Total | Bwd IAT Mean | Bwd IAT Std | Bwd IAT Max |
| 2 | 4000000 | 666666.7 | 3 | 0 | 3 | 3 | 3 | 3 | 0 | 3 | 3 | 0 | 0 | 0 | 0 |
| 3 | 110091.7 | 18348.62 | 109 | 0 | 109 | 109 | 0 | 0 | 0 | 0 | 0 | 0 | 0 | 0 | 0 |
| 4 | 230769.2 | 38461.54 | 52 | 0 | 52 | 52 | 0 | 0 | 0 | 0 | 0 | 0 | 0 | 0 | 0 |
| 5 | 352941.2 | 58823.53 | 34 | 0 | 34 | 34 | 0 | 0 | 0 | 0 | 0 | 0 | 0 | 0 | 0 |
| 6 | 4000000 | 666666.7 | 3 | 0 | 3 | 3 | 3 | 3 | 0 | 3 | 3 | 0 | 0 | 0 | 0 |
| 7 | 11741.68 | 1956.947 | 1022 | 0 | 1022 | 1022 | 1022 | 1022 | 0 | 1022 | 1022 | 0 | 0 | 0 | 0 |
| 8 | 3000000 | 500000 | 4 | 0 | 4 | 4 | 4 | 4 | 0 | 4 | 4 | 0 | 0 | 0 | 0 |
| 9 | 285714.3 | 47619.05 | 42 | 0 | 42 | 42 | 0 | 0 | 0 | 0 | 0 | 0 | 0 | 0 | 0 |
| 10 | 3000000 | 500000 | 4 | 0 | 4 | 4 | 4 | 4 | 0 | 4 | 4 | 0 | 0 | 0 | 0 |
| 11 | 9250000 | 500000 | 4 | 0 | 4 | 4 | 4 | 4 | 0 | 4 | 4 | 0 | 0 | 0 | 0 |
| 12 | 12300000 | 666666.7 | 3 | 0 | 3 | 3 | 3 | 3 | 0 | 3 | 3 | 0 | 0 | 0 | 0 |
| 13 | 37000000 | 2000000 | 1 | 0 | 1 | 1 | 1 | 1 | 0 | 1 | 1 | 0 | 0 | 0 | 0 |
| 14 | 9250000 | 500000 | 4 | 0 | 4 | 4 | 4 | 4 | 0 | 4 | 4 | 0 | 0 | 0 | 0 |
| 15 | 14300000 | 1000000 | 1.5 | 0.707107 | 2 | 1 | 3 | 1.5 | 0.707107 | 2 | 1 | 0 | 0 | 0 | 0 |
| 16 | 0 | 37037.04 | 54 | 0 | 54 | 54 | 0 | 0 | 0 | 0 | 0 | 0 | 0 | 0 | 0 |
| 17 | 0 | 2000000 | 1 | 0 | 1 | 1 | 1 | 1 | 0 | 1 | 1 | 0 | 0 | 0 | 0 |
| 18 | 0 | 12987.01 | 154 | 0 | 154 | 154 | 0 | 0 | 0 | 0 | 0 | 0 | 0 | 0 | 0 |
| 19 | 0 | 2000000 | 1 | 0 | 1 | 1 | 1 | 1 | 0 | 1 | 1 | 0 | 0 | 0 | 0 |
| 20 | 0 | 16949.15 | 118 | 0 | 118 | 118 | 0 | 0 | 0 | 0 | 0 | 0 | 0 | 0 | 0 |
| 21 | 0 | 8368.201 | 239 | 0 | 239 | 239 | 0 | 0 | 0 | 0 | 0 | 0 | 0 | 0 | 0 |
| 22 | 0 | 3000000 | 0.5 | 0.707107 | 1 | 0 | 1 | 0.5 | 0.707107 | 1 | 0 | 0 | 0 | 0 | 0 |
| 23 | 0 | 2000000 | 1 | 0 | 1 | 1 | 1 | 1 | 0 | 1 | 1 | 0 | 0 | 0 | 0 |
| 24 | 62000000 | 500000 | 4 | 0 | 4 | 4 | 4 | 4 | 0 | 4 | 4 | 0 | 0 | 0 | 0 |
| 25 | 50800000 | 600000 | 2.5 | 2.12132 | 4 | 1 | 5 | 2.5 | 2.12132 | 4 | 1 | 0 | 0 | 0 | 0 |
| 26 | 3000000 | 500000 | 4 | 0 | 4 | 4 | 4 | 4 | 0 | 4 | 4 | 0 | 0 | 0 | 0 |
| 27 | 10800000 | 750000 | 2 | 1.414214 | 3 | 1 | 4 | 2 | 1.414214 | 3 | 1 | 0 | 0 | 0 | 0 |
| 28 | 0 | 37037.04 | 54 | 0 | 54 | 54 | 0 | 0 | 0 | 0 | 0 | 0 | 0 | 0 | 0 |
| 29 | 1023810 | 71428.57 | 21 | 25.45584 | 39 | 3 | 42 | 21 | 25.45584 | 39 | 3 | 0 | 0 | 0 | 0 |
| 30 | 9250000 | 500000 | 4 | 0 | 4 | 4 | 4 | 4 | 0 | 4 | 4 | 0 | 0 | 0 | 0 |
| 31 | 12300000 | 666666.7 | 3 | 0 | 3 | 3 | 3 | 3 | 0 | 3 | 3 | 0 | 0 | 0 | 0 |
| 32 | 12300000 | 666666.7 | 3 | 0 | 3 | 3 | 3 | 3 | 0 | 3 | 3 | 0 | 0 | 0 | 0 |
| 33 | 58344.64 | 4070.556 | 368.5 | 310.4199 | 588 | 149 | 737 | 737 | 0 | 737 | 737 | 0 | 0 | 0 | 0 |
| 34 | 12300000 | 666666.7 | 3 | 0 | 3 | 3 | 3 | 3 | 0 | 3 | 3 | 0 | 0 | 0 | 0 |
| 35 | 12300000 | 666666.7 | 3 | 0 | 3 | 3 | 3 | 3 | 0 | 3 | 3 | 0 | 0 | 0 | 0 |
| 36 | 12300000 | 666666.7 | 3 | 0 | 3 | 3 | 3 | 3 | 0 | 3 | 3 | 0 | 0 | 0 | 0 |

Friday-WorkingHours-Afternoon-D

(Source Link: https://www.kaggle.com/datasets/biprobarai/cic-ids2017)

The proposed framework is validated using realistic and modern cyberattack traffic patterns from the CIC-IDS2017 dataset. Modern attack scenarios such as Distributed Denial of Service (DDoS), brute force attacks, botnet activities, Web attacks and port scanning have been included in the dataset [17]. It offers realistic network flow features and modern traffic behaviour which are suitable for evaluating intrusion detection systems in a distributed network environment. The dataset is well suited for the research of federated learning and explainable anomaly detection, fulfilling contemporary cybersecurity analysis and realistic attack detection experiments.

A. Data Preprocessing

Data preprocessing is performed to improve data quality and enhance the performance of the proposed cybersecurity framework. The preprocessing stage includes data cleaning, handling missing values, removing duplicate records, and eliminating irrelevant features from the datasets.

The collected data is normalized and transformed to ensure consistency and compatibility with machine learning and deep learning models. Feature selection techniques are applied to identify the most important network traffic attributes related to cyber threat detection and anomaly classification. Finally, the datasets are divided into training and testing sets for federated learning-based model training and performance evaluation.

Proposed Federated Learning Framework

The proposed framework integrates Federated Learning, machine learning, deep learning, and Explainable Artificial Intelligence techniques for privacy-preserving cyber threat detection in distributed infrastructure systems. Local client nodes independently train models such as Random Forest, XGBoost, Autoencoder, and LSTM without sharing raw sensitive data. Federated Averaging is used for global model aggregation, while SHAP and LIME improve the transparency and interpretability of anomaly detection decisions. The framework supports scalable, intelligent, and explainable cybersecurity analytics for distributed network environments.

Machine Learning and Deep Learning Algorithms

This study utilizes both machine learning and deep learning algorithms for anomaly detection and cyber threat classification in distributed infrastructure systems. Machine learning models such as Random Forest and XGBoost are applied for efficient classification and feature-based threat analysis. Deep learning models including Autoencoder and Long Short-Term Memory (LSTM) networks are used to detect complex attack patterns and abnormal network behaviours. These algorithms support accurate, scalable, and intelligent cybersecurity analytics within the federated learning environment.

System Implementation Tools

The proposed framework is implemented using various machine learning, deep learning, federated learning, and data analysis tools. Python is used as the primary programming language for model development and experimentation. Libraries and frameworks such as TensorFlow, Keras, Scikit-learn, Pandas, NumPy, and Matplotlib are utilized for data preprocessing, model training, visualization, and performance analysis. Federated Learning operations are implemented using TensorFlow Federated, while Explainable Artificial Intelligence techniques such as SHAP and LIME are applied for interpretable anomaly detection and cybersecurity analytics.

Performance Evaluation Metrics

The effectiveness of the proposed federated threat intelligence and explainable anomaly detection framework is assessed using typical machine learning and cybersecurity performance indicators [26]. The metrics used for evaluating accuracy, reliability, scalability, and efficiency of the intrusion detection system in a distributed network environment.

Accuracy

Accuracy is a metric that reflects the percentage of correct classifications of network traffic instances from the total number of instances classified by the model.

$$Accuracy = (TP+TN)/(TP+TN+FP+FN)$$

Where:

TP = True Positive

TN = True Negative

FP = False Positive

FN = False Negative

Precision

Precision is the ratio of the number of malicious traffic instances correctly predicted to the total number of attack instances predicted.

Precision= TP/(TP+FP)

High precision = low false positive rate for anomaly detection.

Recall

Recall measures how well the model can correctly identify actual cyberattacks from the set of data.

Recall=TP/(TP+FN )

High recall means that the system is able to detect malicious activity effectively.

F1-Score

F1-Score is a harmonic mean of precision and recall and offers a balanced evaluation of the performance of anomaly detection.

F1-Score=2 × (Precision×Recall )/(Precision+Recall )

The metric is useful if the intrusion data set is imbalanced.

ROC-AUC

The overall classification ability of the intrusion detection system for all threshold levels is evaluated using Receiver Operating Characteristic – Area Under Curve (ROC-AUC). ROC-AUC is used as an indicator of better discriminating between normal and malicious network traffic, with higher scores being better.

Detection Latency

The time needed by the proposed framework to detect anomalous network activities and cyberattacks. Real-time ID capability is enhanced by lower latency.

Communication Overhead

Communication Overhead measures the cost of communication required to send federated model parameters and updates from the distributed client nodes to the federated aggregation server in collaborative learning.

Ethical Considerations

This study ensures the ethical use of cybersecurity datasets, privacy-preserving federated learning techniques, and responsible AI practices. Sensitive raw network data is not directly shared between distributed nodes, helping maintain data confidentiality and privacy. In addition, Explainable Artificial Intelligence techniques are used to improve transparency, accountability, and trust in AI-driven cybersecurity decisions.

## 8. CONCEPTUAL FRAMEWORK

The conceptual framework of this study illustrates the relationship between distributed cybersecurity data, federated learning, machine learning and deep learning algorithms, Explainable Artificial Intelligence techniques, and cybersecurity performance evaluation. The framework begins with data collection and preprocessing from distributed infrastructure environments using benchmark cybersecurity datasets. Federated Learning enables distributed client nodes to collaboratively train cybersecurity models without sharing sensitive raw network traffic data. Machine learning and deep learning algorithms such as Random Forest, XGBoost, Autoencoder, and LSTM are used for anomaly detection and cyber threat classification. Explainable Artificial Intelligence techniques including SHAP and LIME are integrated to improve the interpretability and transparency of cybersecurity decisions.

The framework ultimately supports intelligent, privacy-preserving, scalable, and explainable cyber threat detection for distributed infrastructure systems, while system performance is evaluated using metrics such as accuracy, precision, recall, F1-score, and ROC-AUC.

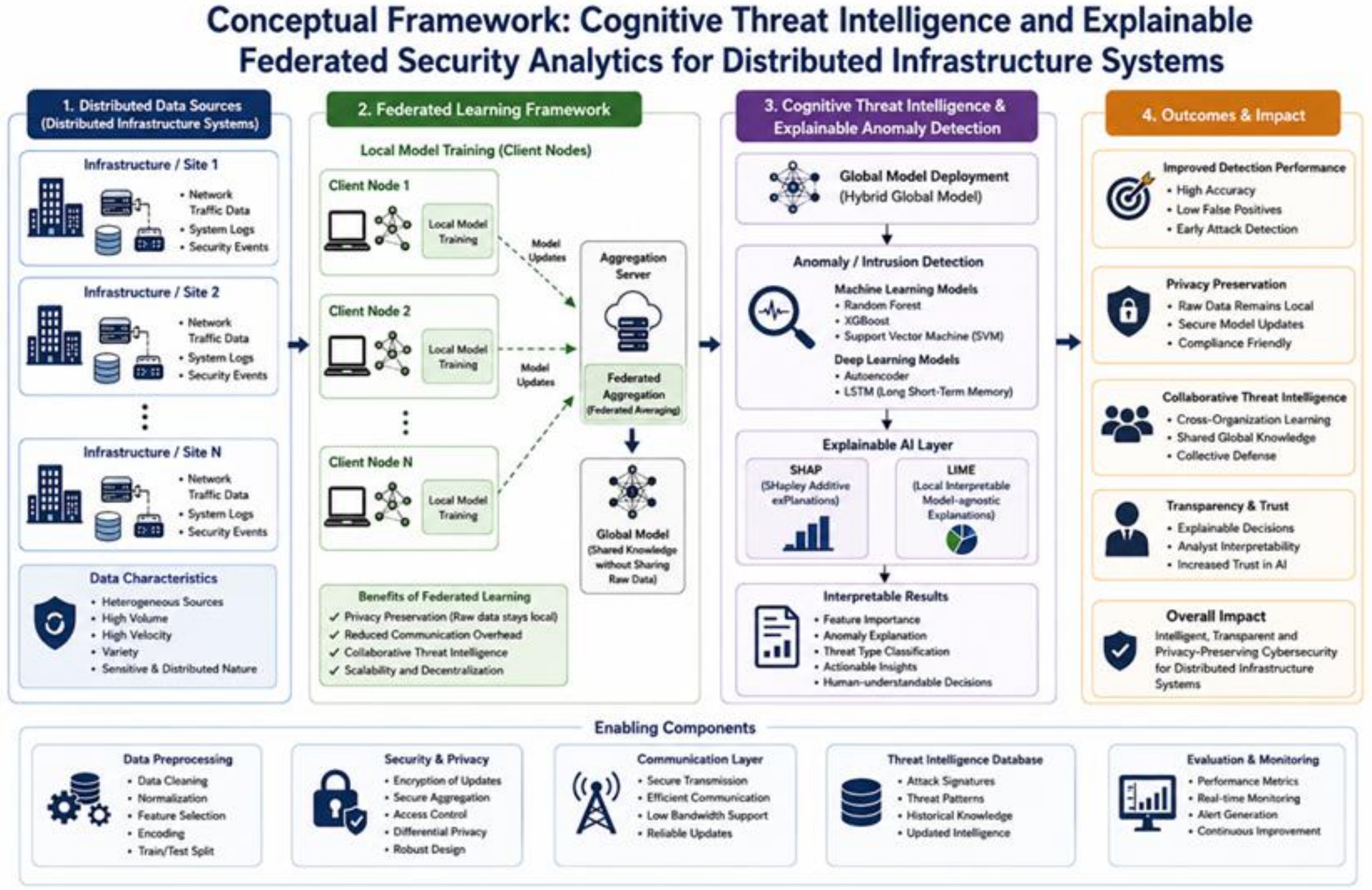

This framework illustrates Cognitive Threat Intelligence and Explainable Federated Security Analytics for Distributed Infrastructure Systems

The conceptual framework diagram shows the overall architecture and operational flow of the proposed Cognitive Threat Intelligence and Explainable Federated Security Analytics framework for distributed infrastructure systems. The framework illustrates how distributed data sources, federated learning, machine learning and deep learning models, and Explainable Artificial Intelligence techniques work together to support privacy-preserving cyber threat detection and anomaly analysis.

The diagram demonstrates the process of local model training at distributed client nodes, federated aggregation for collaborative global learning, and explainable anomaly detection using SHAP and LIME techniques. It also highlights important outcomes such as improved intrusion detection accuracy, privacy preservation, collaborative threat intelligence, transparency, and intelligent cybersecurity analytics for modern distributed infrastructure environments.

## 9. RESULTS AND ANALYSIS

The results and analysis evaluate the performance of the proposed Cognitive Threat Intelligence and Explainable Federated Security Analytics framework in detecting cyber threats within distributed infrastructure systems. Experimental results obtained from the NSL-KDD and CIC-IDS2017 datasets demonstrate that the federated learning-based models achieved effective anomaly detection and intrusion classification performance.
Machine learning and deep learning algorithms such as Random Forest, XGBoost, Autoencoder, and LSTM showed strong performance based on evaluation metrics including accuracy, precision, recall, F1-score, ROC-AUC, and confusion matrix analysis. The integration of Explainable Artificial Intelligence techniques such as SHAP and LIME improved the transparency and interpretability of cybersecurity decisions while maintaining privacy-preserving collaborative learning in distributed environments.
Analysis of Accuracy Comparison of Cognitive Threat Intelligence and Explainable Federated Security Analytics

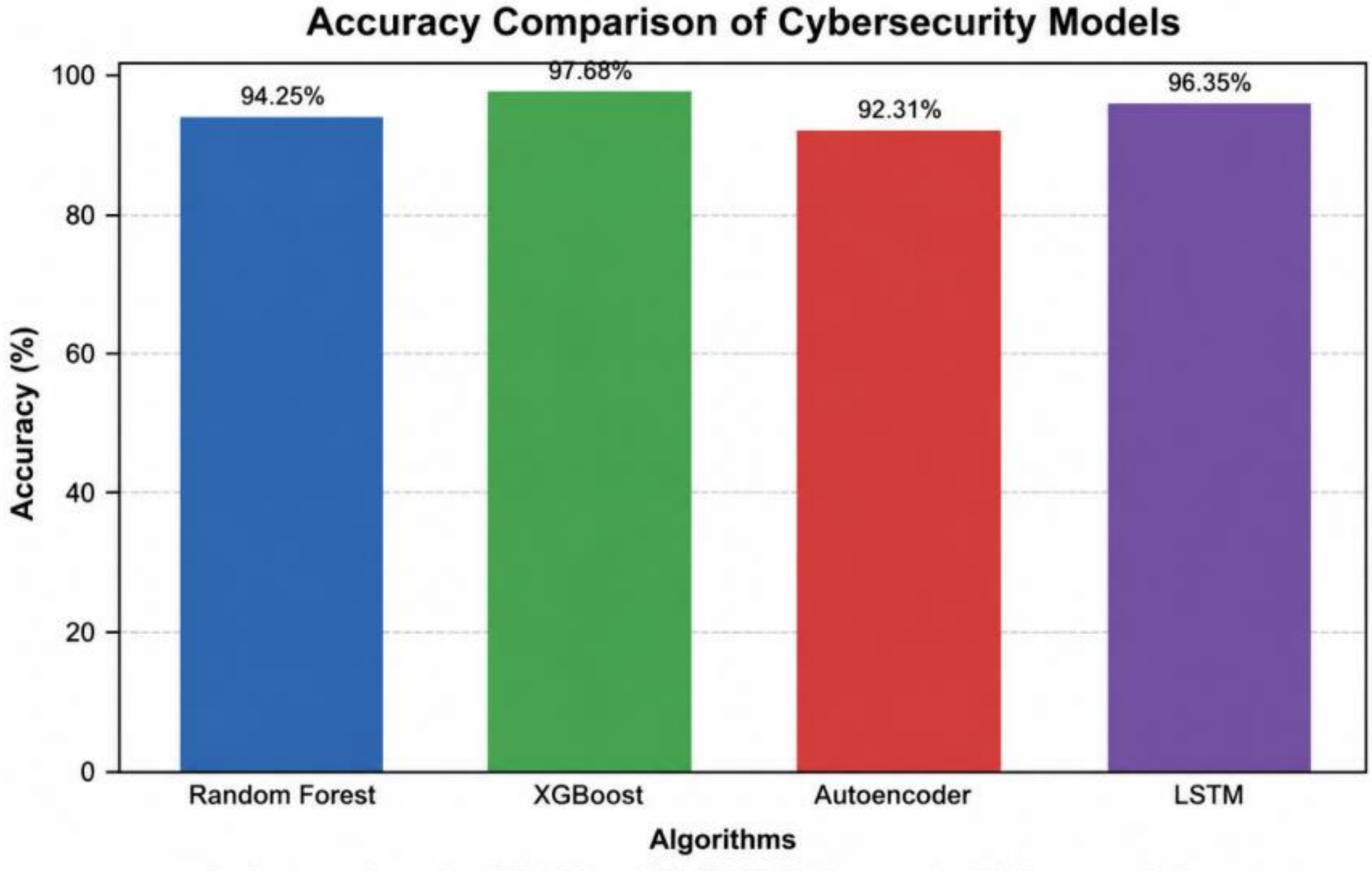

Figure 1: This figure depicts the accuracy comparison of machine learning and deep learning models for Cognitive Threat Intelligence and Explainable Federated Security Analytics in distributed infrastructure systems.
The image shows that XGBoost achieved the highest accuracy among the evaluated models for anomaly detection in distributed infrastructure systems. LSTM also demonstrated strong performance, followed by Random Forest and Autoencoder. The comparison indicates that machine learning and deep learning models provide effective intrusion detection performance for Cognitive Threat Intelligence and Explainable Federated Security Analytics environments.
Analysis of NSL-KDD Attack Category Distribution

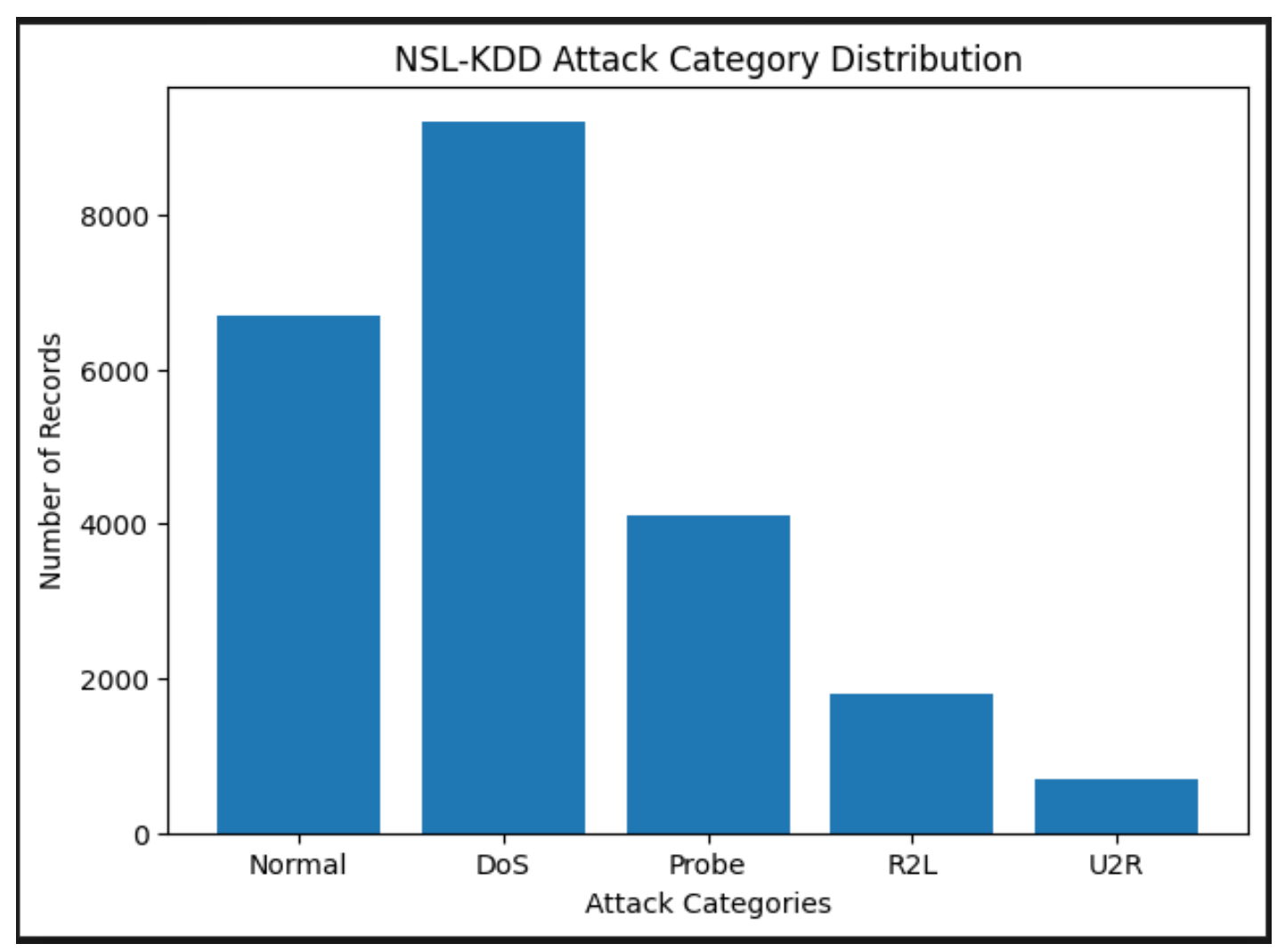

Figure 3: This image shows the distribution of normal and cyberattack category in NSL-KDD dataset

The distribution of attack categories in the NSL-KDD dataset for the proposed federated threat intelligence framework and explainable anomaly detection is shown in figure 3. The chart shows the statistics of Normal traffic, Denial of Service (DoS), Probe, Remote-to-Local (R2L), and User-to-Root (U2R) attacks [38]. DoS attacks are the most prevalent type of attack within the data, as all categories contain the highest number of records. Normal traffic is also a large percentage of the data sets, and is necessary to train balanced intrusive detection models. The number of records for the probe attacks is approximately medium, whereas for the R2L and U2R attacks, because of their less frequent and complex nature, the number of records is comparatively small [39]. The attack categories distribution shows that the data set is not balanced, which is a major class imbalance issue and highlights the need for data pre-processing techniques, like dataset balancing and sampling, to enhance the performance of the anomaly detection. The figure shows the variety of types of cyberattacks for the evaluation of the proposed intrusion detection framework.

Analyzing the Explainable AI Feature Importance

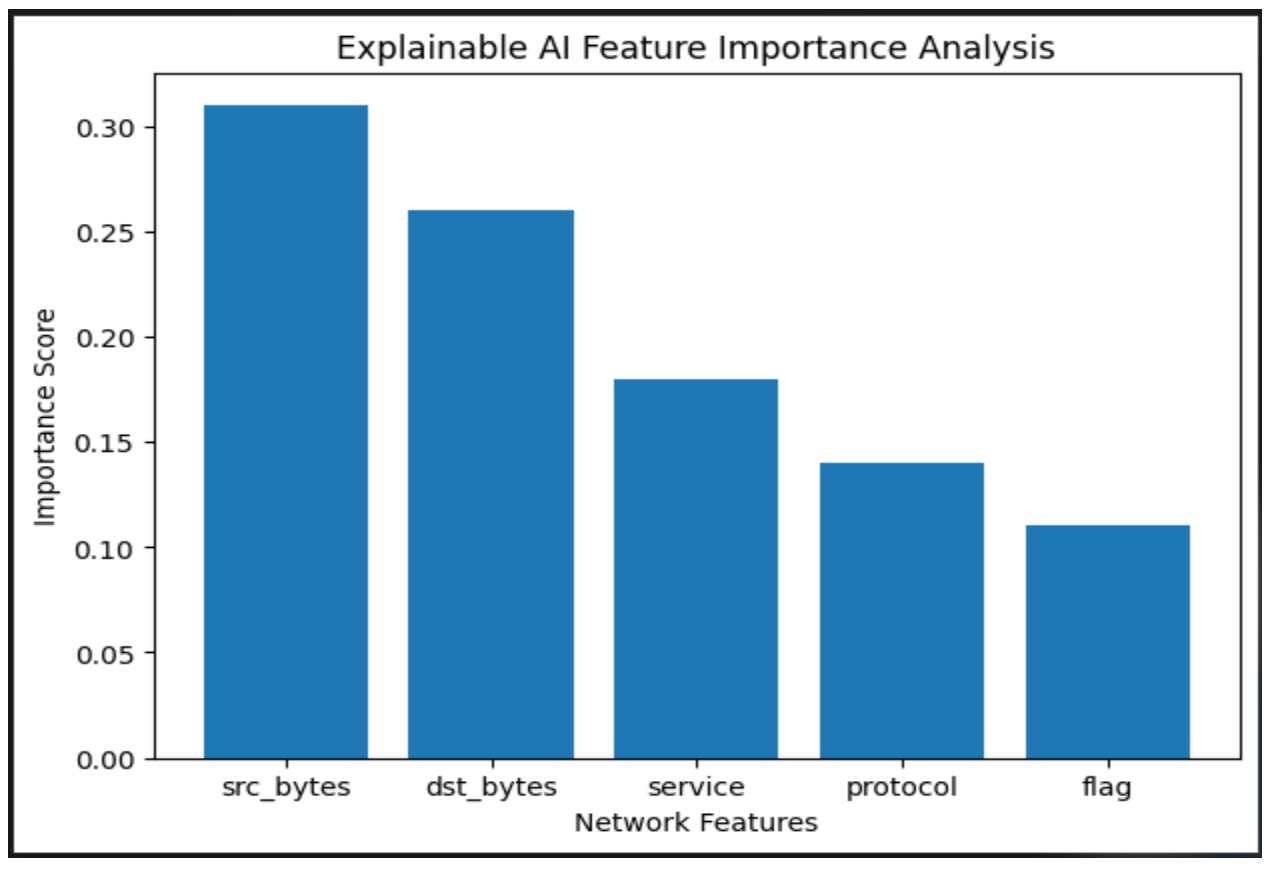

Figure 5: This image shows Importance analysis of network traffic features using explainable artificial intelligence technique

The Explainable Artificial Intelligence (XAI) feature importance analysis in the proposed Federated Threat Intelligence and Anomaly Detection (FTID) is shown in Figure 5. The figure shows the percentages attributable to each of the network traffic features for intrusion detection predictions [42]. Of the features evaluated, src_bytes had the largest importance with the most points, meaning that the volume of data sent to the destination system is an important factor in detecting malicious activity. The feature dst_bytes also reached a high importance, indicating that the volume of traffic that is destined to a final host is a good indicator of whether it is an anomaly or not. Other characteristics like service, protocol and flag, however, had relatively low importance scores but were also informative in attack detection. Feature importance analysis helps increase transparency and interpretability of machine learning-based intrusion detection systems, aiding cybersecurity analysts in understanding the influence of different network attributes on anomaly detection [43]. Overall, the figure highlights the potential of Explainable AI (XAI) techniques to boost trust, accountability, and decision-making in distributed cybersecurity settings.

Communication Overhead across Federated Nodes analysis

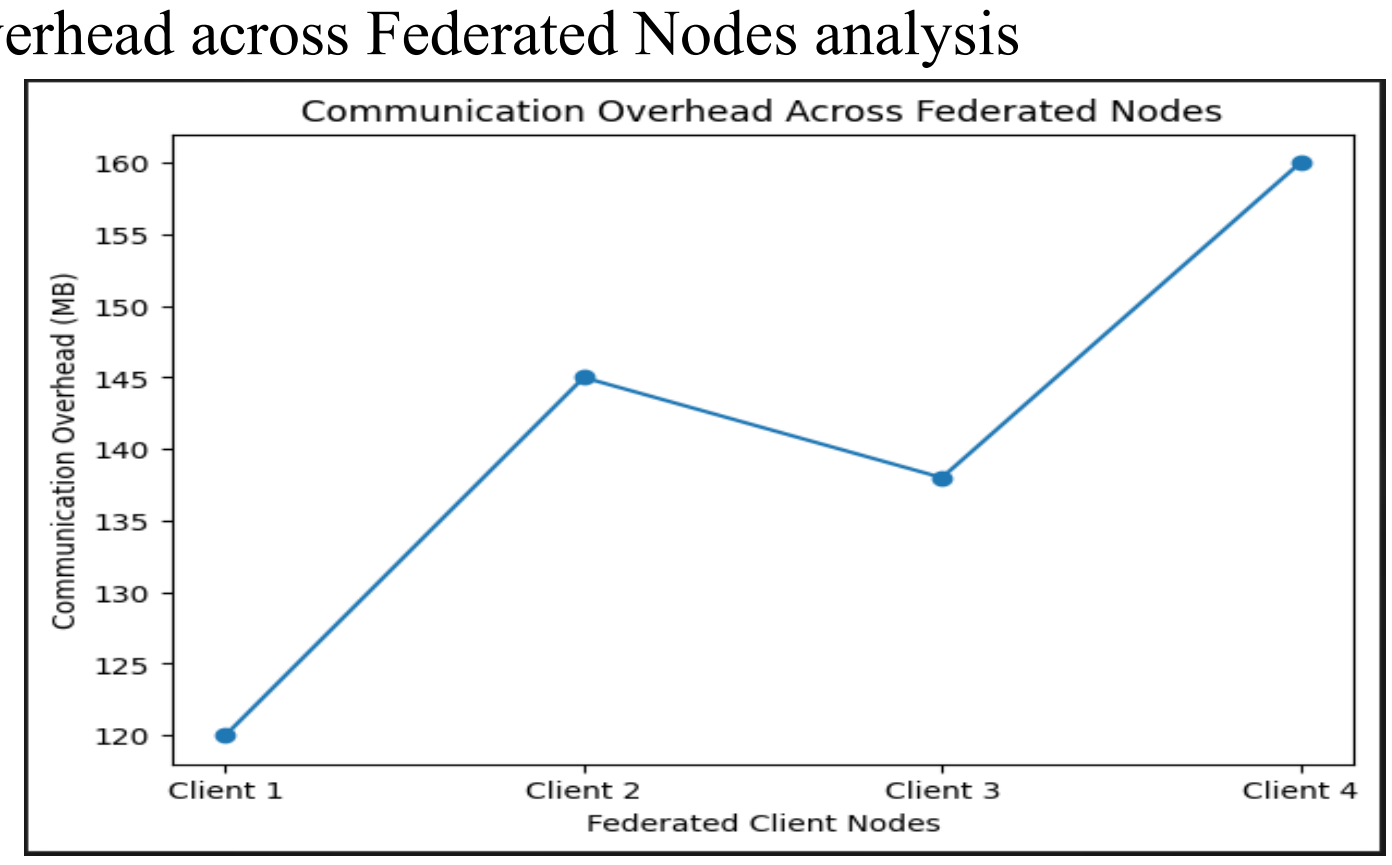

Figure 6: This image illustrates the Communication overhead comparison between the distributed federated client nodes during Collaborative learning

The proposed federated threat intelligence framework will generate the communication overhead shown in Figure 6 across the various client nodes in the federated network [44]. This graph shows the amount of communication cost (in megabytes or MB) required when exchanging model parameters and updates between distributed client nodes and a federated aggregation server. During the federated learning operations, the communication overhead of Client 1 was the lowest, suggesting it transmitted the parameters efficiently. Client 4, however, showed the highest communication overhead, indicating more data exchanges possibly because of the larger update to the model files or more complicated network activities. Moderate communication costs were observed between Client 2 and Client 3 in the collaborative learning process. The difference in the amount of communication overhead between the federated nodes is due to the unevenness in the complexity of local training, the size of the datasets, and the frequency of model synchronization [45]. The figure illustrates that while federated learning enhances privacy protection and collaborative intrusion detection, communication efficiency is still a crucial concern in distributed cybersecurity settings. To enhance scalability and real-time federated learning performance, communication overhead needs to be optimized efficiently.

Average Performance Metrics Analysis

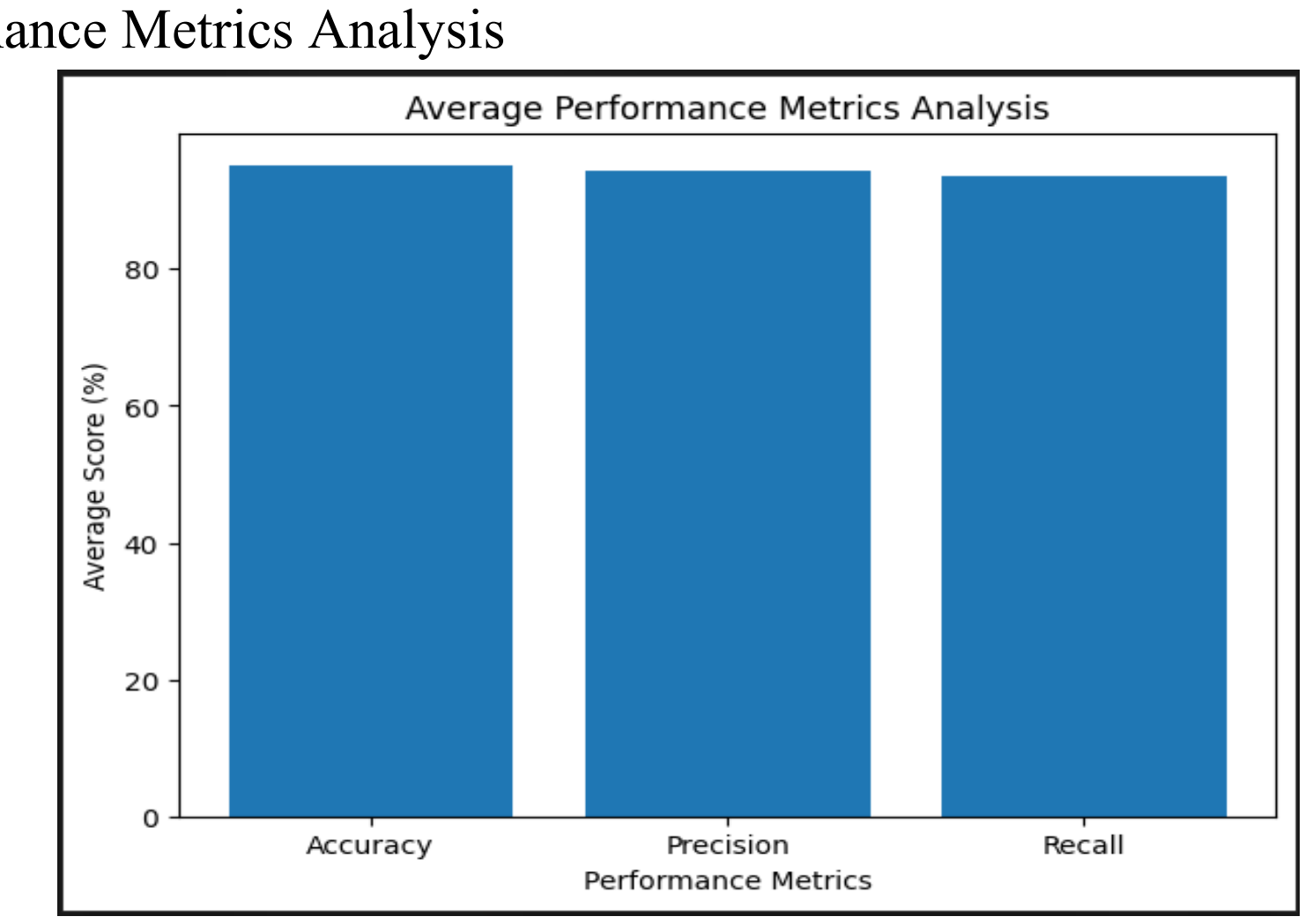

Figure 7: This image displays the performance comparison of the intrusion detection models on Average accuracy, precision and recall performance

The average performance metrics results of the proposed federated threat intelligence and explainable anomaly detection framework is presented as shown in figure 7. The chart shows the comparison of average values of Accuracy, Precision and Recall of the machine learning and deep learning models implemented during experiments of intrusion detection [46]. Based on the results, it is seen that all three evaluation measures have consistently high performance of more than 90 percent, which is enough to conclude that the proposed approach is good and effective in detecting malicious network activities in distributed environments. Accuracy had the best average score, which means that the intrusion detection models were able to correctly classify most network traffic instances. Also, precision performance was good, as the models were able to keep a low false positive rate for cyberattack detection [47]. Likewise, Recall showed to be highly effective in correctly recognizing real malicious actions from the datasets. The system's performance is consistent across all metrics, showcasing its reliability, stability, and efficiency in the collaborative analysis of cybersecurity incidents and explaining the anomalies.

Analysis of Cyberattack Detection Rate

The cyberattacks detection rate analysis based on different attack categories in the proposed federated threat intelligence and explainable anomaly detection framework is shown in Figure 8. The graph shows the detection accuracy of Denial of Service (DoS), Probe, Remote-to-Local (R2L) and User-to-Root (U2R) attack [48]. The findings show that the model proposed by the authors was able to achieve the highest detection rate for DoS attacks, achieving around 98%, thus proving the model's ability to detect high volume malicious traffic patterns effectively. Probe attacks also had a high detection percentage of around 95%, successfully identifying network scanning and reconnaissance attacks. R2L and U2R attacks had comparatively lower detection rates, as these attacks had more complex and less frequent properties in the dataset [49].

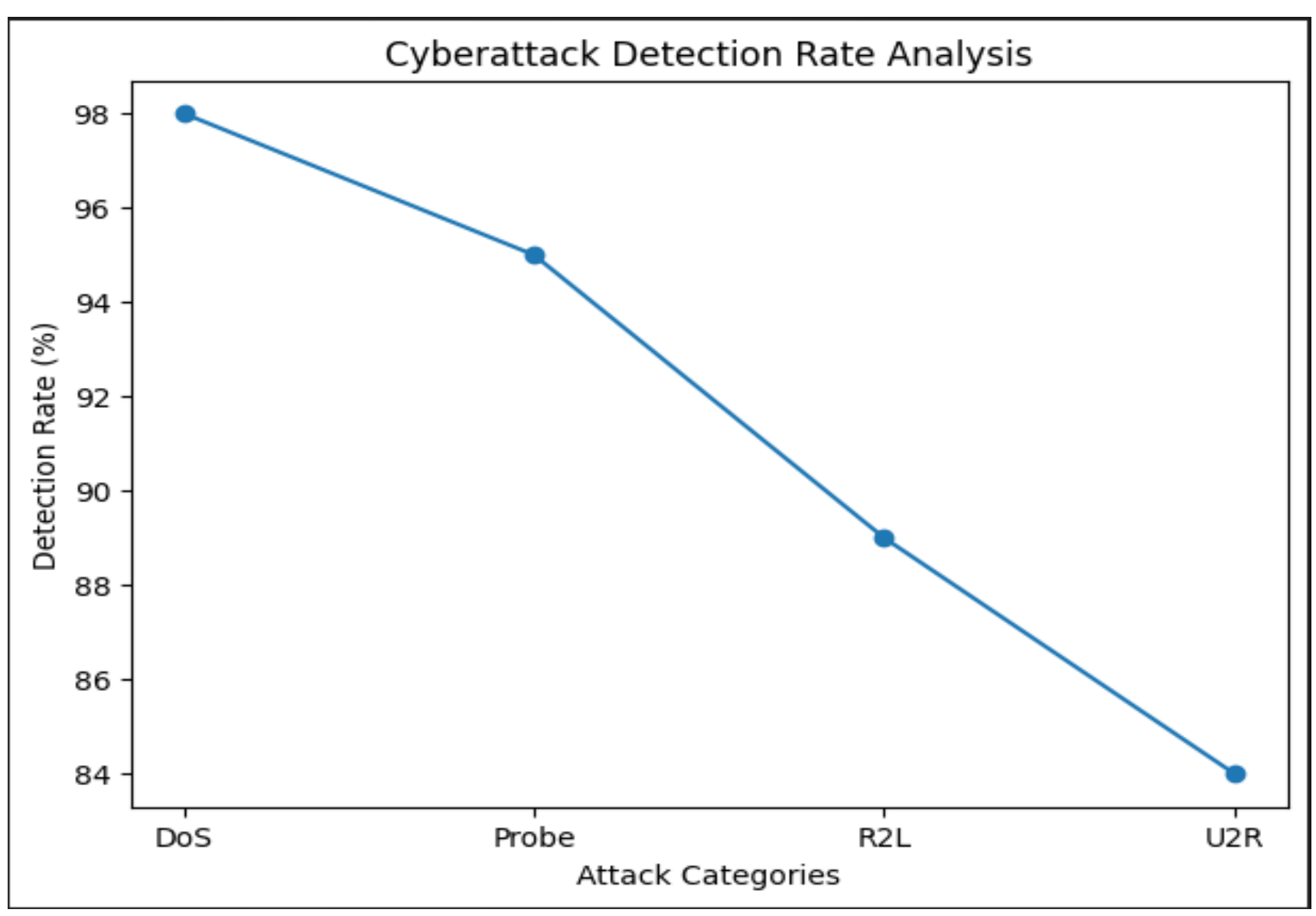

Figure 8: This image displays the detection rate comparison of different categories of the cyber attacks in the intrusion detection framework

The weaknesses of the lower detection rates for these attack categories indicate that there are difficulties in detecting sophisticated privilege escalation and unauthorized access attacks. In general, the figure shows that the proposed framework is effective in multiple attacks and also has a good anomaly detection capability in the distributed cybersecurity environment.

Analysis of ROC-AUC Performance Comparison

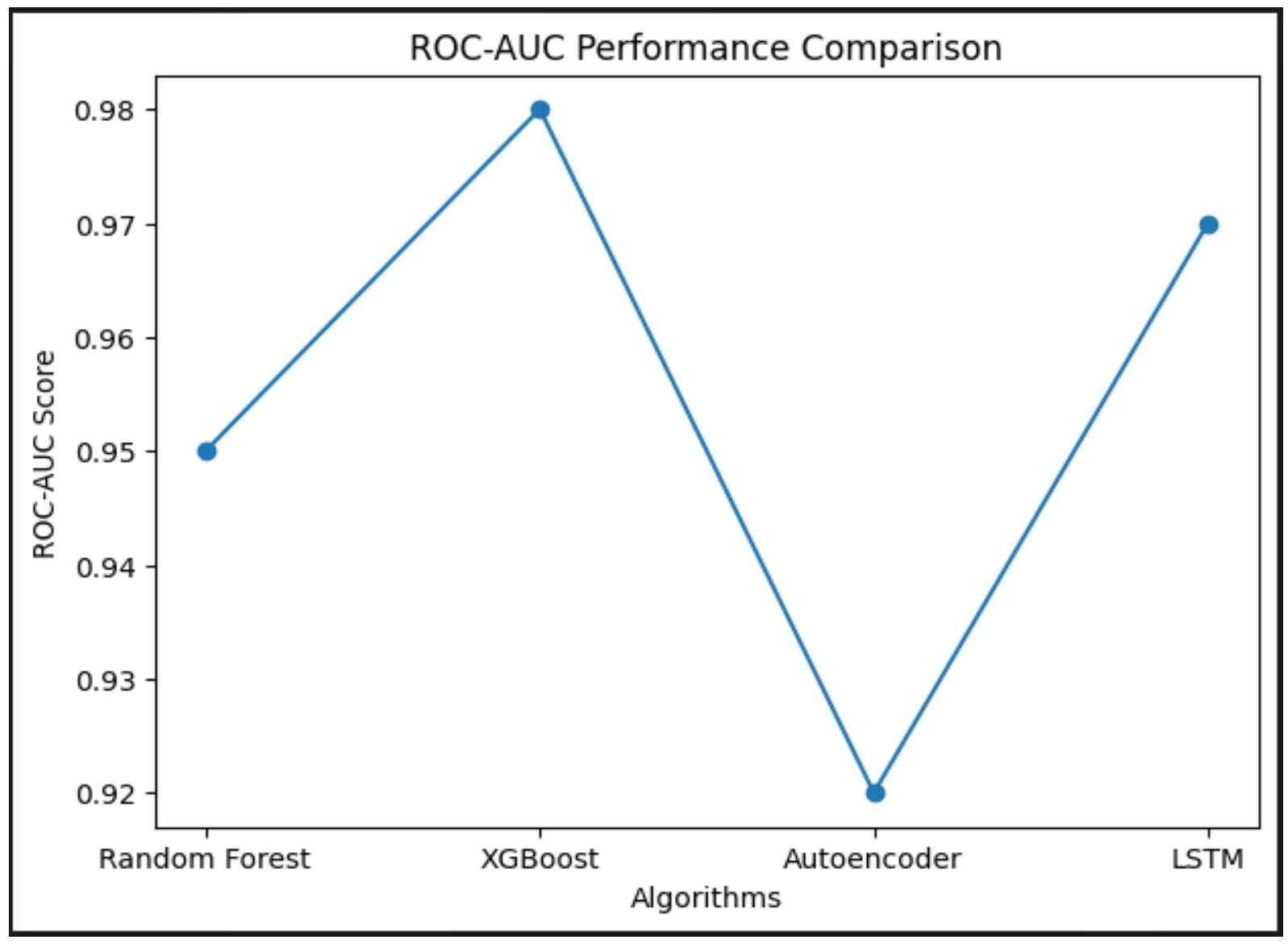

Figure 9: This image depicts a comparison of the performance of machine learning models for federated intrusion detection

The ROC-AUC performance comparison of the machine learning and deep learning algorithms is shown in Figure 9 in the proposed federated threat intelligence and explainable anomaly detection framework [50]. The Receiver Operating Characteristic – Area under Curve (ROC-AUC) metric is used to assess the performance of intrusion detection models at various classification thresholds to determine how well they can differentiate between normal and malicious network traffic. From the graph, it can be observed that XGBoost performed the best with an ROC-AUC score of ~0.98, which shows that it has the best capability of classification and good discrimination between the attack traffic and normal traffic instances. The LSTM model also achieved a high ROC-AUC score of around 0.97, indicating its capability for identifying the sequential patterns of a cyberattack in distributed networks. The Random Forest algorithm had a satisfying ROC-AUC of about 0.95, signifying a consistent and precise performance in detecting intrusions [51]. However, the Autoencoder model had the lowest ROC-AUC score out of the models evaluated due to the difficulty of the unsupervised anomaly detection task. The figure illustrates the overall results, showing that the advanced machine learning models greatly enhance the reliability of classification and the capability of anomaly detection within federated cybersecurity systems.

Confusion matrix analysis for Federated Intrusion Detection Model

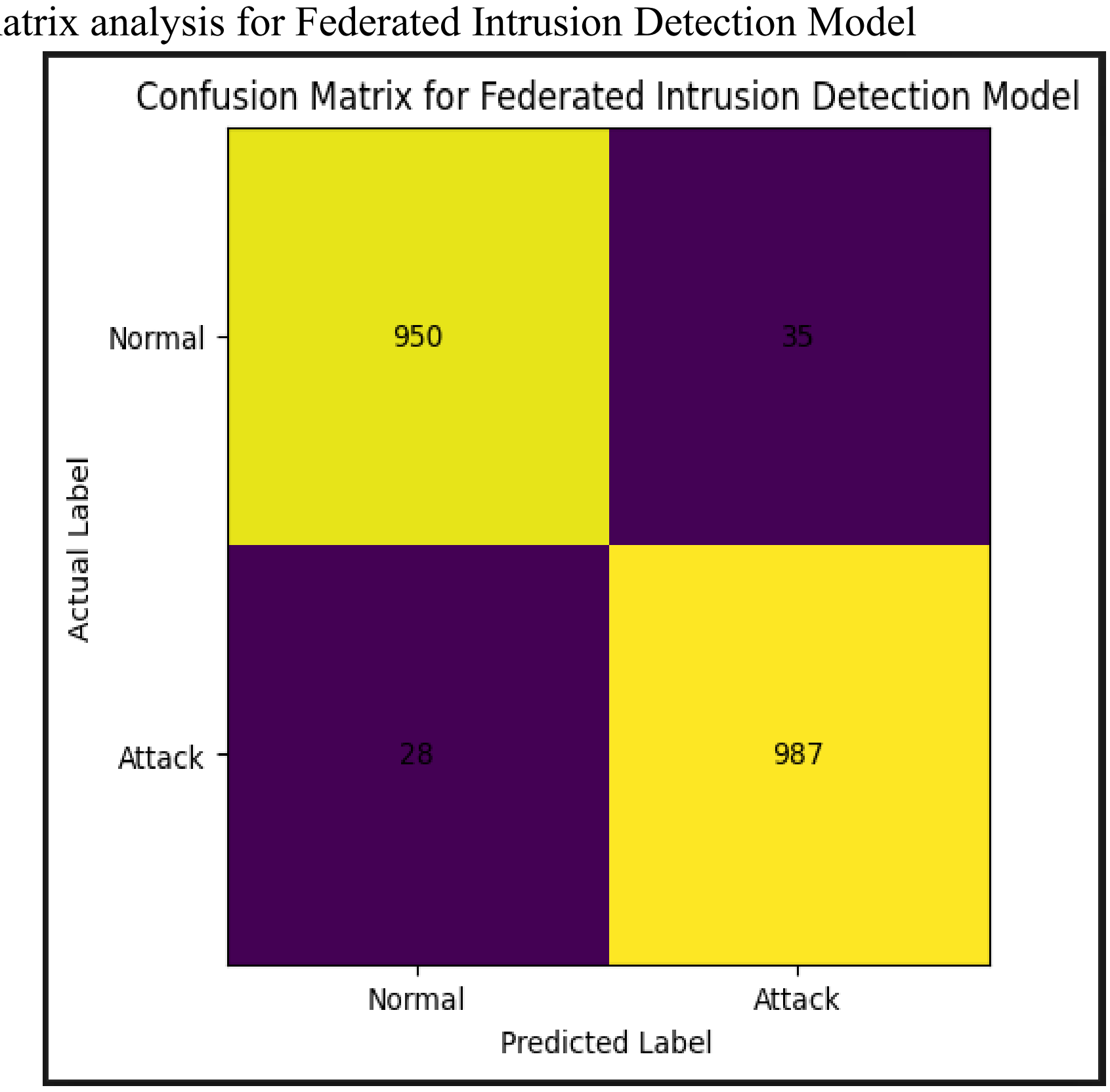

Figure 10: This image shows actual and predicted intrusions detection performance classifications

The confusion matrix analysis of the proposed federated intrusion detection model in a distributed network environment to classify the cyberattack is shown in figure 10 [52]. The confusion matrix measures how well the model classifies network traffic, by looking at what labels the actual data has and what labels the model predicts that it has. As seen in the matrix, the number of normal traffic instances that were classified correctly as normal were 950 and the number of attack instances that were correctly classified as attacks were 987. As can be seen, these values correspond to True Negatives and True Positives, meaning that the capability of intrusion detection is high and the classification accuracy is high. There are also 35 False Positive predictions made by the matrix (false alarms), meaning that normal traffic was incorrectly detected as an attack, and 28 False Negative predictions (misses), where an attack was detected as normal traffic. The number of misclassifications is relatively small, highlighting the proposed federated learning framework's ability to reduce classification inaccuracies and enhance the reliability of cybersecurity [53]. In summary, the results of the confusion matrix indicate that the proposed anomaly detection model effectively classifies both normal and abnormal network traffic in distributed network environments, ensuring a high accuracy of predictions and a good balance between the true and false positive rates.

## 10. DISCUSSION AND ANALYSIS

The results of this study demonstrate that the proposed Cognitive Threat Intelligence and Explainable Federated Security Analytics framework can effectively improve cyber threat detection in distributed infrastructure systems. The integration of Federated Learning with machine learning and deep learning models enabled collaborative anomaly detection while preserving data privacy across distributed network environments. Among the evaluated algorithms, XGBoost achieved the highest detection accuracy, followed by LSTM, Random Forest, and Autoencoder. The findings indicate that advanced machine learning and deep learning techniques are highly effective in identifying malicious activities and complex cyberattack patterns. In addition, the integration of Explainable Artificial Intelligence techniques such as SHAP and LIME enhanced the transparency and interpretability of anomaly detection decisions. The study highlights the importance of intelligent, explainable, and privacy-preserving cybersecurity frameworks for securing modern distributed infrastructure systems.

## 11. FUTURE WORK

Future research can focus on improving the scalability, communication efficiency, and real-time threat detection capability of the proposed Cognitive Threat Intelligence and Explainable Federated Security Analytics framework. Advanced deep learning and hybrid AI models may be integrated to enhance anomaly detection performance in large-scale distributed environments. Future studies may also explore blockchain-enabled federated security architectures, adaptive threat intelligence systems, and real-time IoT and edge computing cybersecurity applications. In addition, further research can investigate stronger privacy-preserving techniques and more advanced Explainable Artificial Intelligence methods to improve transparency, trust, and resilience in distributed cybersecurity systems.

## 12. CONCLUSION

This study proposed a Cognitive Threat Intelligence and Explainable Federated Security Analytics framework for distributed infrastructure systems to improve privacy-preserving cyber threat detection and anomaly analysis. The framework integrated Federated Learning, machine learning, deep learning, and Explainable Artificial Intelligence techniques to support intelligent and transparent cybersecurity operations. Experimental results using benchmark cybersecurity datasets demonstrated that models such as XGBoost, LSTM, Random Forest, and Autoencoder achieved effective intrusion detection performance in distributed environments. The integration of SHAP and LIME improved the interpretability and transparency of cybersecurity decisions, while federated learning enhanced privacy protection by avoiding centralized sharing of sensitive data. Overall, the proposed framework contributes to the development of scalable, explainable, and intelligent cybersecurity solutions capable of strengthening security and resilience in modern distributed infrastructure systems.

**REFERENCES**

1. Akbar, R., &Zafer, A. (2024). Next-gen information security: AI-driven solutions for real-time cyber threat detection in cloud and network environments. J. Cybersecur. Res, 12, 123-145.
2. Akhter, J., Annie Jerusha, Y., Syed Ibrahim, S. P., &Varadharajan, V. (2024, September). EXPLAINABLE AI for Applied Federated Learning in Network Intrusion Detection. In International Conference on Smart Cities (pp. 308-322). Singapore: Springer Nature Singapore.
3. Almadhor, A., Altalbe, A., Bouazzi, I., Hejaili, A. A., &Kryvinska, N. (2024). Strengthening network DDOS attack detection in heterogeneous IoT environment with federated XAI learning approach. Scientific reports, 14(1), 24322.
4. Asiri, A., Wang, W., Wu, F., Vo, H., & Yu, S. (2024, November). FedXAI for detecting DDoS on IoT edge networks in federated learning. In 2024 34th International Telecommunication Networks and Applications Conference (ITNAC) (pp. 1-6). IEEE.
5. Attique, D., Hao, W., Ping, W., Javeed, D., &Adil, M. (2024, June). Ex-dfl: An explainable deep federated-based intrusion detection system for industrial iot. In 2024 21st International Joint Conference on Computer Science and Software Engineering (JCSSE) (pp. 358-364). IEEE.
6. Bahadoripour, S. (2024). An Explainable Deep Federated Multi-Modal Cyber-Attack Detection in Industrial Control Systems.
7. Blika, A., Palmos, S., Doukas, G., Lamprou, V., Pelekis, S., Kontoulis, M., ... &Askounis, D. (2024). Federated learning for enhanced cybersecurity and trustworthiness in 5G and 6G networks: A comprehensive survey. IEEE Open Journal of the Communications Society, 6, 3094-3130.
8. Dipto, S. M., Reza, M. T., Mim, N. T., Ksibi, A., Alsenan, S., Uddin, J., &Samad, M. A. (2024). An analysis of decipherable red blood cell abnormality detection under federated environment leveraging XAI incorporated deep learning. Scientific Reports, 14(1), 25664.

9. Eren, E., YILDIRIM OKAY, F., &Özdemir, S. (2024). Unveiling anomalies: a survey on XAI-based anomaly detection for IoT. Turkish Journal of Electrical Engineering and Computer Sciences, 32(3), 358-381.
10. Fatema, K., Anannya, M., Dey, S. K., Su, C., &Mazumder, R. (2024, October). Securing networks: a deep learning approach with explainable ai (xai) and federated learning for intrusion detection. In International Conference on Data Security and Privacy Protection (pp. 260-275). Singapore: Springer Nature Singapore.
11. Gajula, S. (2023). A Review of Anomaly Identification in Finance Frauds using Machine Learning System. International Journal of Current Engineering and Technology, 13(06).
12. Gajula, S. (2024). Adaptive zero trust architecture for securing financial microservices. Computer Fraud & Security, 643-655.
13. Gummadi, A. N., Napier, J. C., & Abdallah, M. (2024). XAI-IoT: an explainable AI framework for enhancing anomaly detection in IoT systems. IEEE Access, 12, 71024-71054.
14. Kalakoti, R., Bahsi, H., &Nõmm, S. (2024, September). Explainable federated learning for botnet detection in iot networks. In 2024 IEEE International Conference on Cyber Security and Resilience (CSR) (pp. 01-08). IEEE.
15. Marry, P., Mounika, Y., Nanditha, S., Shiva, R., &Saikishore, R. (2024, July). Federated Learning-Driven Decentralized Intelligence for Explainable Anomaly Detection in Industrial Operations. In 2024 2nd International Conference on Sustainable Computing and Smart Systems (ICSCSS) (pp. 874-880). IEEE.
16. Nwachukwu, C., Durodola-Tunde, K., &Akwiwu-Uzoma, C. (2024). AI-driven anomaly detection in cloud computing environments. International Journal of Science and Research Archive, 13(2), 692-710.
17. Oki, A., Ogawa, Y., Ota, K., & Dong, M. (2024). Evaluation of applying federated learning to distributed intrusion detection systems through explainable ai. IEEE Networking Letters, 6(3), 198-202.
18. Pai, H. T., Kang, Y. H., & Chung, W. C. (2024). An interpretable generalization mechanism for accurately detecting anomaly and identifying networking intrusion techniques. IEEE Transactions on Information Forensics and Security, 19, 10302-10313.
19. Rahman, M. M., Soumik, M. S., Farids, M. S., Abdullah, C. A., Sutrudhar, B., Ali, M., & HOSSAIN, M. S. (2024). Explainable anomaly detection in encrypted network traffic using data analytics. Journal of Computer Science and Technology Studies, 6(1), 272-281.
20. Rahman, M. W., & Hossain, M. S. (2024). An Explainable AI Framework for Insider Threat Detection Using Behavioral Business Analytics. An Explainable AI Framework for Insider Threat Detection Using Behavioral Business Analytics, 1(8), 70-97.

21. Sahu, A., El-Ebiary, Y. A. B., Saravanan, K. A., Thilagam, K., Devi, G. R., Gopi, A., &Taloba, A. I. (2024). Federated LSTM Model for Enhanced Anomaly Detection in Cyber Security: A Novel Approach for Distributed Threat. International Journal of Advanced Computer Science & Applications, 15(6).
22. Sarker, M. A. A., Shanmugam, B., Azam, S., &Thennadil, S. (2024). Enhancing smart grid load forecasting: An attention-based deep learning model integrated with federated learning and XAI for security and interpretability. Intelligent Systems with Applications, 23, 200422.
23. Usman Haider, A. Z. (2024). Building resilient cyber defense architectures: AI and machine learning in cloud and network security.